\documentclass[aapm,graphicx,groupedaddress,superscriptaddress]{revtex4-1}
\draft
\setcitestyle{super}
\usepackage[sort&compress]{natbib}
\usepackage{amsfonts}
\usepackage{amssymb}
\usepackage{amsmath,epsfig}
\usepackage{subfigure}
\usepackage{fmtcount}
\usepackage{array}
\usepackage{rotating}
\usepackage{fmtcount}
\usepackage{mathbbol}
\usepackage{dsfont}
\usepackage{placeins}
\usepackage{amsmath}
\usepackage{booktabs,xcolor,siunitx}
\usepackage{slashbox}
\usepackage{url}
\usepackage[caption=false]{subfig}
\usepackage{textcomp}
\usepackage{times}
\usepackage{textcomp}

\usepackage{multirow}

\usepackage{threeparttable}
\usepackage{nomencl}
\usepackage{soul}
\usepackage{color}
\usepackage{algorithm}
\usepackage{algorithmic}
\usepackage{subfigure}
\usepackage{setspace}
\usepackage{algorithm}
\usepackage{algorithmic}

\makenomenclature

\usepackage{makeidx}
\makeindex 

\begin{document}
\def\vector{\underline}
\def\matrix{}
\renewcommand{\vec}[1]{\mathbf{#1}}
\newcommand{\mat}[1]{\mathbf{#1}}

\title{Low Rank Magnetic Resonance Fingerprinting}

\author{Gal Mazor}
\affiliation{Department of Electrical Engineering, Technion - Israel Institue of Technology, Israel}
\author{Lior Weizman}
\affiliation{Department of Electrical Engineering, Technion - Israel Institue of Technology, Israel}
\author{Assaf Tal}
\affiliation{Department of Chemical Physics, Weizmann Institute of Science, Rehovot, Israel}
\author{Yonina C. Eldar}
\address{Department of Electrical Engineering, Technion - Israel Institue of Technology, Israel}

\begin{abstract}
{\bf Purpose:}  Magnetic Resonance Fingerprinting (MRF) is a relatively new approach that provides quantitative MRI measures using randomized acquisition. 
Extraction of physical quantitative tissue parameters is performed off-line, without the need of patient presence, based on acquisition with varying parameters and a dictionary generated according
to the Bloch equation simulations. 
MRF uses hundreds of radio frequency (RF) excitation pulses for acquisition, and therefore a high under-sampling ratio in the sampling domain (k-space) is required for reasonable scanning time. This under-sampling causes spatial artifacts that hamper the ability to accurately estimate the tissue's quantitative values. In this work, we introduce a new approach for quantitative MRI using MRF, called magnetic resonance Fingerprinting with LOw Rank (FLOR).

{\bf Methods:} We exploit the low rank property of the concatenated temporal imaging contrasts, on top of the fact that the MRF signal is sparsely represented in the generated dictionary domain. We present an iterative scheme that consists of a gradient step followed by a low rank projection using the singular value decomposition.

{\bf Results:} Experimental results consist of retrospective sampling, that allows comparison to a well defined reference, and prospective sampling that shows the performance of FLOR for a real-data sampling scenario.
Both experiments demonstrate improved parameter accuracy compared to other compressed-sensing and low-rank based methods for MRF at 5\% and 9\% sampling ratios, for the retrospective and prospective experiments, respectively.

{\bf Conclusions:}
We have shown through retrospective and prospective experiments that by exploiting the low rank nature of the MRF signal, FLOR recovers the MRF temporal undersampled images and provides more accurate parameter maps compared to previous iterative methods. 
\end{abstract}



\maketitle

\newpage

\section{Introduction}
\label{sec:intro}

Quantitative Magnetic Resonance Imaging (QMRI) is widely used to measure tissue's intrinsic spin parameters such as the spin-lattice magnetic relaxation time (T1) and the spin-spin magnetic relaxation time (T2) \cite{ehses2013ir}. Since tissue relaxation times vary 
in disease, QMRI enables the diagnosis of different pathologies, including multiple sclerosis (MS), Alzheimer, Parkinson, epilepsy and cancer \cite{baselice2016bayesian,antonini1993t2,baselice2014optimal,haley2004shortening,mariappan1988proton,roebuck2009carr}. In addition, the knowledge of tissue relaxation times allows generation of many clinical MR imaging contrasts (such as FLAIR and STIR) off-line, and saves a significant amount of scanning time. 


Despite the advantages of QMRI, clinical MRI today mainly consists of weighted images. Values in weighted MR imaging are given in arbitrary units, since the signal strength is influenced by both intrinsic parameters (such as relaxation times and concentration of hydrogen atoms) and non-intrinsic ones. Non-intrinsic parameters include transmit and receive coils sensitivities, patient position in the scanner, vendor based scanner specific parameters, and local temperature. Weighted MRI images therefore lack quantitative information and as a result, different materials may exhibit similar or identical gray level values. In addition, weighted MRI contrast values vary between different follow-up scans of the same patient. This fact may impair disease monitoring, if based solely on those images. To date, weighted MRI scans are  more common than QMRI in the clinic, due to the extremely long times often associated with QMRI using conventional techniques \cite{liu1989calculation,homer1985driven,crooijmans2011finite}.

A plethora of methods have been proposed for QMRI. Earlier approaches are based on a series of spin echo (SE) or inversion recovery (IR) images with varying repetition times (TR) and echo times (TE) to evaluate each magnetic parameter (T1 or T2) separately. After acquisition, the curve of intensities for each pixel is matched to the expected magnetic signal, representing the appropriate magnetic tissue parameters \cite{liu1989calculation}. Accelerated methods for QMRI consist of the popular driven equilibrium single pulse observation of T1 (DESPOT1) \cite{homer1985driven} or T2 (DESPOT2) \cite{crooijmans2011finite} and the IR TrueFISP for simultaneous recovery of T1 and T2 quantitative maps \cite{schmitt2004inversion,ehses2013ir}. 
Both techniques do not require long waiting times between excitations to reach an equilibrium state, and therefore they are significantly faster. Later works shortened the acquisition time required by those methods by under-sampling the data in both spatial and temporal domains \cite{doneva2010compressed,zhao2015accelerated,petzschner2011fast,huang2012t2,velikina2013accelerating}. 
However, obtaining accurate and high resolution QMRI in a reasonable clinical scanning time is still very challenging. 

An approach for QMRI called magnetic resonance fingerprinting (MRF) has drawn increased attention in the last few years \cite{ma2013magnetic}. MRF uses pseudo-randomized acquisitions to generate many different imaging contrasts, acquired at a high under-sampling ratio.   
It exploits the different acquisition parameters over time to produce a temporal signature, a ``fingerprint", for each material under investigation. By matching this unique signature to a pre-generated set of simulated patterns, the quantitative parameters can be extracted off-line. This approach saves valuable scan time compared to previous methods for accelerated QMRI, demonstrating promising efficient and reliable results.

MRF utilizes the fact that each tissue responds differently to a given quasi-random pulse sequence. By varying the acquisition parameters (e.g. repetition time (TR), echo time (TE), and radio frequency flip angle (FA)), 
unique signals are generated from different tissues. After acquisition, a pattern recognition algorithm is used to match the acquired signal from each voxel to an entry from a dictionary of possible tissue candidates. The dictionary entries are created by simulating the tissue's response to the sequence for a range of T1 and T2 parameter values, using the Bloch equations. The resulting dictionary contains the temporal signatures of various simulated materials, given the pseudo-random pulse sequence. The quantitative parameters, such as the tissue's T1 and T2 relaxation times, can be retrieved from the data by matching the signature acquired to the most correlated entry in the dictionary.

\begin{figure}
\begin{center}
\vspace{2mm}
\includegraphics[scale=0.4, clip=true]{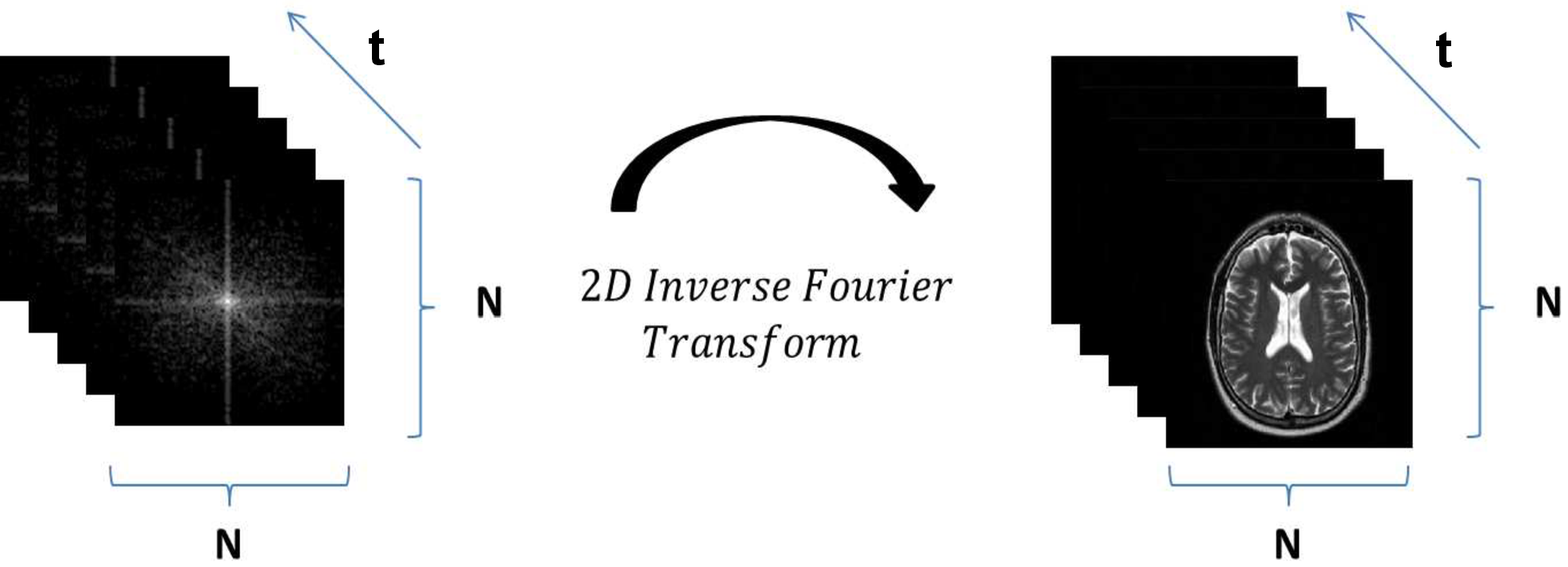} 
\includegraphics[scale=0.4, clip=true]{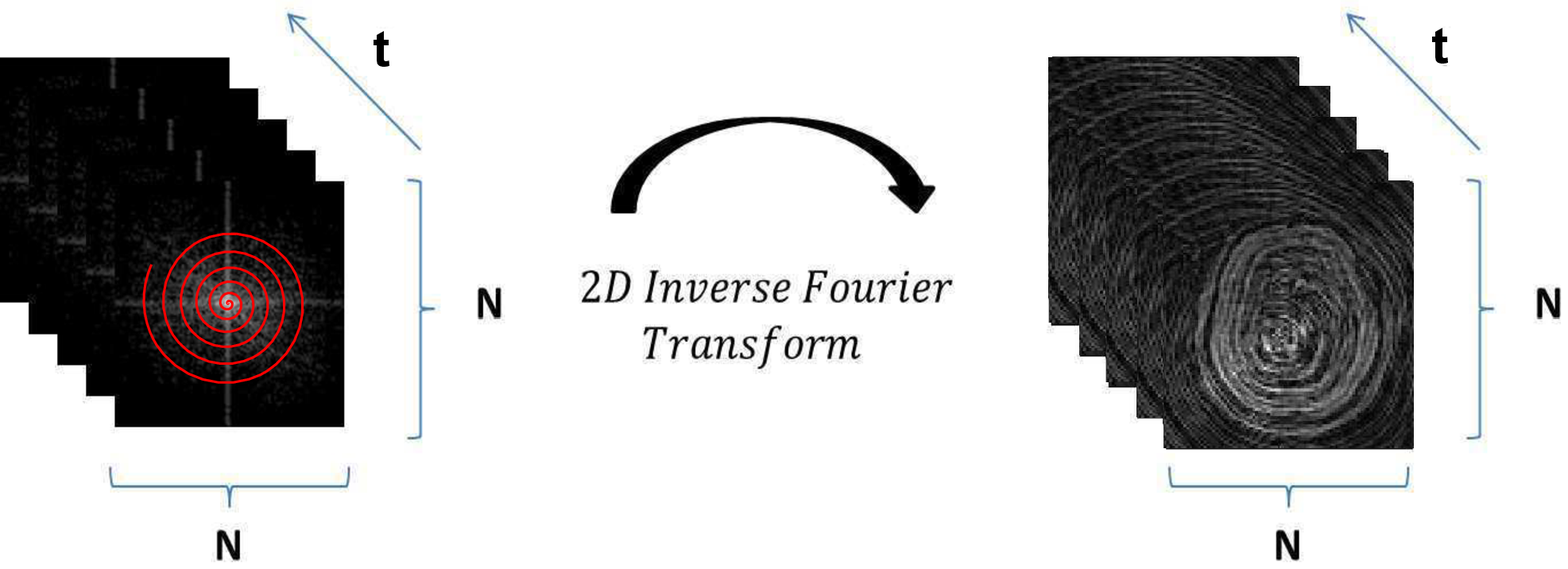} 
\caption{Illustration of fully sampled (left) vs. spiral trajectory under-sampled (right) k-spaces and their corresponding reconstructed images using direct inverse Fourier transform. In MRF, we acquire many under-sampled images over time. When reconstructing by NUFFT, the under-sampled data is blurred and contains aliasing artifacts.}
\label{2DiF}
\vspace{2mm}
\end{center}
\end{figure}

\begin{figure}
\begin{center}
\begin{turn}{90}\parbox{2cm}{\hspace{0mm} voxel intensity}\end{turn}\hspace{1mm}\includegraphics[width = 14cm,height=20mm,trim=2.7cm 1.1cm 1cm 0cm, clip=true]{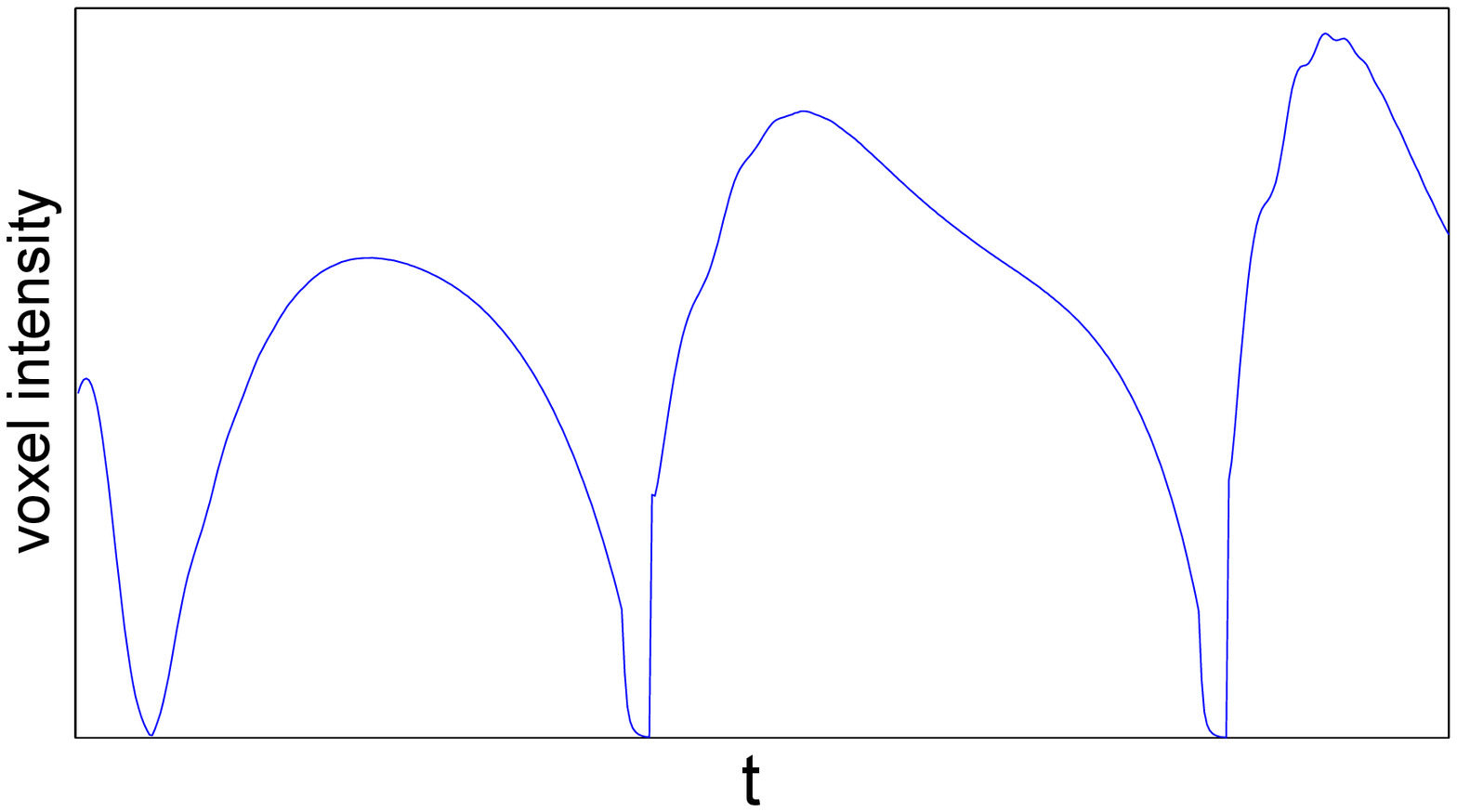}\\
{\hspace*{0cm} t}\\
\begin{turn}{90}\parbox{2cm}{\hspace{0mm} voxel intensity}\end{turn}\hspace{1mm}\includegraphics[width = 14cm,height=20mm,trim=2.7cm 1.1cm 1cm 0cm, clip=true]{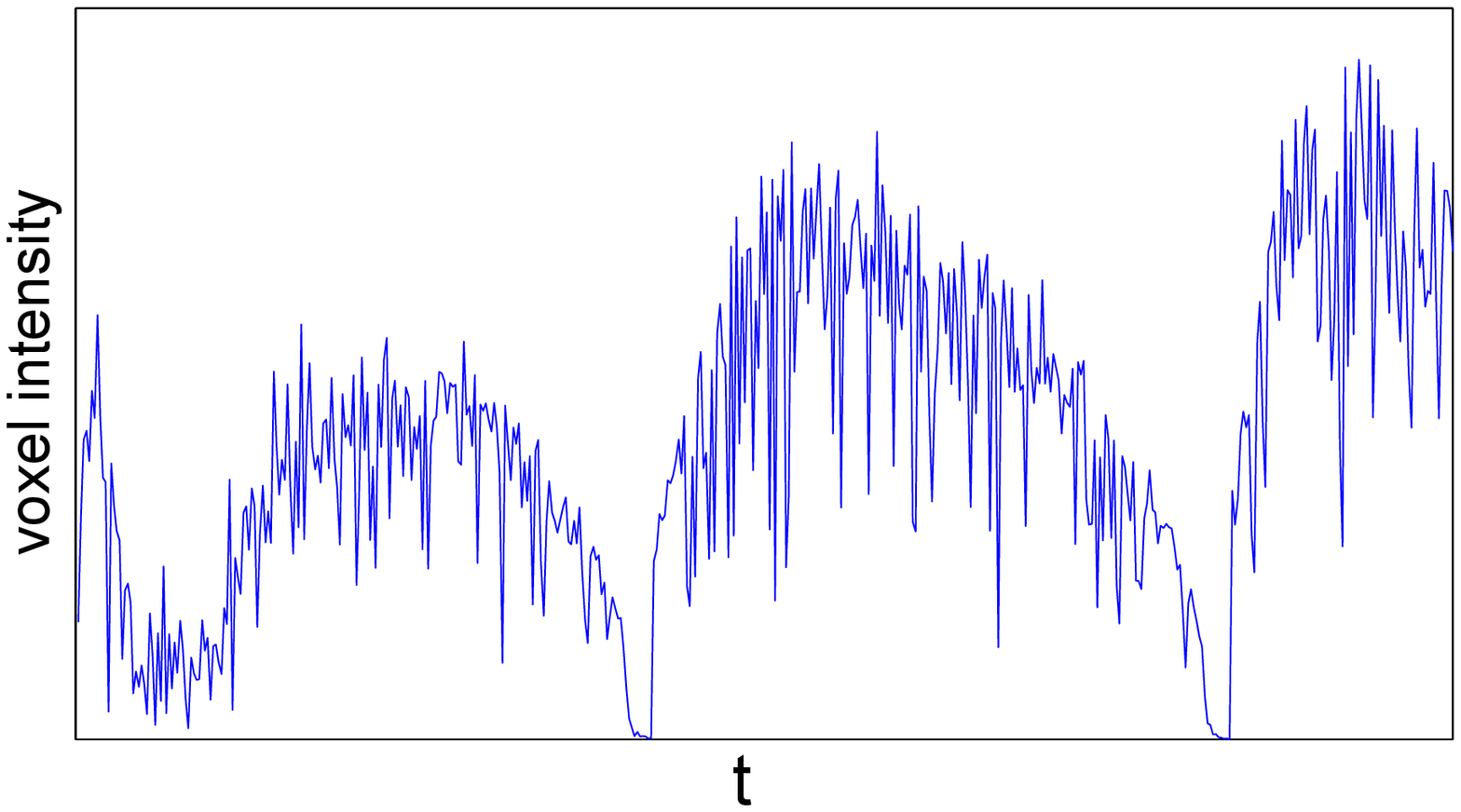}\\
{\hspace*{0cm} t}\\
\caption{Illustration of a single brain voxel's temporal signature acquired with an MRF approach based on the FISP sequence. Fully sampled (top) vs. noisy and under-sampled (bottom) at 15\% sampling ratio.}
\label{PI}
\vspace{2mm}
\end{center}
\end{figure}

In MRI, data is acquired in the Fourier domain of the spatial image (a.k.a. k-space). The acquisition time of a high resolution, single contrast 3D MRI lasts a substantial amount of time. Since MRF is based on rapid acquisition of hundreds of different contrasts, severe under-sampling is performed in k-space to obtain the temporal resolution required for MRF. Figure~\ref{2DiF} demonstrates the effect of fully sampled versus under-sampled data, acquired with spiral trajectories and recovered using the inverse non uniform fast Fourier transform (NUFFT) \cite{fessler2003nonuniform}. It can be seen that the under-sampled data is blurred and introduces aliasing artifacts. Figure~\ref{PI} illustrates the noise and under-sampling artifacts of a representative brain voxel intensity as function of time, where the data is acquired with an MRF sequence based on fast imaging with steady state precession (FISP) \cite{jiang2015mr}. Clearly, under-sampling also introduces a substantial level of noise in the time domain. In addition, MRF uses a dictionary with discrete values, while QMRI values are continuous. This leads to quantization error, depending on the values represented in the dictionary. 

While in the original MRF paper \cite{ma2013magnetic} these imaging artifacts are not handled explicitly, 
recent works have implemented advanced reconstruction techniques to overcome under-sampling artifacts. 
Approaches based on exploiting the sparsity of the signal in some transform domain in a compressed sensing (CS) \cite{eldar2012compressed,eldar2015sampling} framework are examined by Davies et al. \cite{davies2014compressed} and Wang et al. \cite{wang2016magnetic}. Zhao et al. \cite{zhao2016maximum} formulated MRF as a maximum likelihood (ML) problem, and developed an iterative reconstruction approach for each time point. While these techniques showed improved results compared to the original MRF method, they do not exploit the temporal similarity between adjacent time-points, which is intrinsic to the dynamic acquisition used in MRF. 

A common approach to exploit redundancy exists in dynamic MRI, based on modeling the acquired data as low-rank. This modelling was successfully applied for various dynamic MRI applications, such as cardiac imaging \cite{feng2013highly} and functional MRI \cite{chiew2015k}. In the context of MRF, early works use low-rank MRF to compress the dictionary for faster reconstruction \cite{mcgivney2014svd,cline2016model}. This saves reconstruction time, but does not necessarily improve the quality of the reconstructed maps or the acquisition time. The first introduction of a low-rank constraint for improved reconstruction in MRF was proposed by Zhao et al. \cite{zhao2015model,zhao2016model} followed by a sub-space constrained low-rank approach introduced by us \cite{mazor2016low}. Extensions of these ideas include adding a sparse term to the low-rank-based reconstruction \cite{liao2016acceleration} (a.k.a robust PCA \cite{candes2011robust}) and representing the data as low-rank in the k-space domain \cite{doneva2016low,doneva2017matrix}. Recently, a few approaches that utilize prior knowledge of the dictionary together with a low-rank constraint have been published. Zhao et al. \cite{zhao2017improved} presented an efficient algorithm that performs a singular value decomposition (SVD) on the dictionary and embeds the right singular vectors in the solution, to obtain better estimation of the temporal signatures. A similar approach was presented by Assl\"ander et al. \cite{asslander2017low}, who embed the left singular vectors in the solution. These methods show that exploiting the redundancy via a low-rank based solution improves the results compared to a sparsity approach. However, the obtained reconstructed maps still suffer from quantization error, due to the nature of a matched-filter based solution that matches a single dictionary atom to a single pixel. In addition, most of these methods are based on a fixed rank, set in advance, which may be difficult to determine in advance.

In this paper we extend our initial idea presented in our conference paper \cite{mazor2016low} and enforce a low-rank constraint in the image domain together with constraining the solution to the dictionary subspace. In particular, we exploit the low-rank property of the temporal MRF domain, via an iterative scheme that consists of a gradient step followed by a projection onto the subspace spanned by the dictionary elements in order to constrain the structure of the tissue behaviour simulated in the dictionary. The estimated images are then decomposed using SVD and the singular values are soft-thresholded to reduce the nuclear norm value in every step. Our approach, called magnetic resonance Fingerprinting with LOw Rank constraint (FLOR), incorporates three main advantages that were only partially introduced in previous work:
\begin{itemize}
  \item FLOR formulates the problem as a convex problem. The solution is then rigorously developed based on the incremental subgradient proximal method \cite{nedic2001incremental}. This technique is known to convergence to the global minimum, regardless of the initial starting point.
  \item FLOR is based on a nuclear-norm solution, and does not require fixing the rank in advance. This leads to a solution that adapts the rank according to the nature of the specific dataset.
  \item The subspace constraint in FLOR is not limited to dictionary items, but rather allows a solution that is spanned by the dictionary. This enables better reconstruction of the temporal imaging contrasts. It also allows generation of quantitative parameters that do not necessarily exist in the simulated dictionary, thereby reducing the quantization error of the resulting maps.
\end{itemize}
While there are previous publications that introduce one or two of the advantages pointed above (e.g. Zhao et al. \cite{zhao2017improved} describes a subspace constraint that is not limited to the dictionary items), our work incorporates all of them together in a convenient optimization framework. 
 
 Our reconstruction results are based on sampling with variable density spiral trajectories, using 5\% and 9\% sampling ratios, for retrospective and prospective experiments, respectively. We compare our results to the methods developed by Davies et al. \cite{davies2014compressed} and Zhao \cite{zhao2015model}, and show that FLOR provides quantitative parameter maps with higher accuracy or correspondence to literature compared to those methods. 





This paper is organized as follows. Section \ref{sec:method} describes the MRF problem and provides a review of common reconstruction methods, followed by our low-rank based approach. Section \ref{sec:results} compares our results to previous MRF algorithms, using retrospective and prospective under-sampled MRF data of a human subject. Sections \ref{sec:discussion} and \ref{sec:conclusions} discuss experimental results using retrospective and prospective under-sampled MRF data, followed by conclusions. 

\section{Method}
\label{sec:method}

\newlength\myindent
\setlength\myindent{10em}
\newcommand\bindent{%
  \begingroup
  \setlength{\itemindent}{\myindent}
  \addtolength{\algorithmicindent}{\myindent}
}
\newcommand\eindent{\endgroup}
\renewcommand{\algorithmicrequire}{\textbf{Input:}}
\renewcommand{\algorithmicensure}{\textbf{Output:}}

\subsection{Problem formulation}

MRF data consists of multiple frames, acquired in the image's conjugate Fourier domain (a.k.a k-space), where each frame results from different acquisition parameters. We stack the measurements into a $Q \times L$ matrix $\mathbf{Y}$, where $L$ is the number of frames and $Q$ is the number of k-space samples in each frame. Every column in $\mathbf{Y}$ is an under-sampled Fourier transform of an image frame, $\mathbf{X}_{:,i}$:
\begin{equation}
\mathbf{Y}=[F_u\{\mathbf{X}_{:,1}\},...,F_u\{\mathbf{X}_{:,L}\}]+\mathbf{H}
\end{equation}
where $F_u\{\cdot\}$ denotes an under-sampled 2D Fourier transform and $\mathbf{H}$ denotes a zero mean complex Gaussian noise. The row $\mathbf{X}_{j,:}$ represents the temporal signature of a single pixel (assumed to correspond to a single tissue). The signature depends on the tissue's relaxation times, T1 and T2, and its proton density  (PD), grouped as a row vector:
\begin{equation}
\mathbf{\Theta}_{1}^{j}=[T1^{j},T2^{j},PD^{j}],\quad1\leq j\leq N. 
\label{eq2}
\end{equation}
Each column, $\mathbf{X}_{:,i}$ represents a response image acquired at a single time point with different acquisition parameters, stacked as a column vector: 
\begin{equation}
\mathbf{\Theta}_{2}^{i}=[TR^{i},TE^{i},FA^{i}]^{T},\quad1\leq i\leq L
\label{eq1}
\end{equation}
where TR and TE are the repetition time and time to echo and FA represents
the flip angle of the RF pulse. Therefore, $\mathbf{X}_{j,:}=f(\mathbf{\Theta}_{1}^{j},\mathbf{\Theta}_{2})$, where $f\{\cdot\}$ represents the Bloch equations.
Note that we omit the off resonance parameter (which appeared in $\mathbf{\Theta}_{1}$ in the original MRF paper \cite{ma2013magnetic}), since the sequence used in our retrospective experiments is derived from the FISP sequence, which is insensitive to off resonance effects \cite{jiang2015mr}. 



The goal in MRF is to recover, from the measurements $\mathbf{Y}$, the imaging contrasts $\mathbf{X}$ and the underlying quantitative parameters of each pixel defined in  (\ref{eq2}), under the assumptions that every pixel in the image contains a single type of tissue and that $\mathbf{\Theta}_{{2}}$ is known. 

Recovery is performed by defining a dictionary that consists of simulating the signal generated from $M$ tissues using the Bloch equations (represented as $M$ different combinations of T1 and T2 relaxation times), when the length-$L$ acquisition sequence defined in (\ref{eq1}) is used. 
As a result, we obtain a dictionary $\mathbf{D}$ of dimensions $M \times L$ ($M>L$ as the number of simulated tissues is greater than the sequence length).  The PD is not simulated in the dictionary, as it is the gain used to match the Bloch simulation performed on a single spin to the signal obtained from a pixel containing multiple spins. It can be easily determined after the T1 and T2 maps are known. After successful recovery of $\mathbf{X}$, each row in $\mathbf{X}$ is matched to a single row in the dictionary, and T1 and T2 are estimated as those used to generate the matched dictionary row. Each dictionary signature has its own unique T1 and T2 values stored in a look up table (LUT), represented as the matrix $\mathbf{LUT}$ of dimensions $M\times2$.


\subsection{Previous Methods}

The approach suggested in the original MRF paper \cite{ma2013magnetic} is described in Algorithm \ref{alg1}, and uses matched filtering to match dictionary
items to the acquired data. In the algorithm, $F^{H}\{\cdot\}$ is the 2D inverse NUFFT operator. 
The parameters $k_{j}$ are the matching dictionary indices, $j$ is a spatial index and $i$ is the temporal index, representing the $i$th frame in the acquisition.
The parameter maps are extracted from $\mathbf{LUT}$, which holds the values of T1 and T2 for each $k_{j}$.
This approach does not incorporate sparse based reconstruction, which has been proven to be very successful in MRI applications based on under-sampled data \cite{lustig2008compressed,weizman2015compressed,weizman2016reference}.

\begin{algorithm}[H]
\caption{Original MRF algorithm}
\textbf{Input:}\\
$\quad$ A set of under-sampled k-space images: $\mathbf{Y}$\\
$\quad$ A pre simulated dictionary: $\mathbf{D}$\\
$\quad$ An appropriate look up table: $\mathbf{LUT}$\\
\textbf{Output:}\\
Magnetic parameter maps: $\widehat{T}_{1}, \widehat{T}_{2}, \widehat{PD}$\\
\textbf{Compute for every $i$ and $j$:}\\

\hspace{1cm} $\widehat{\mathbf{X}}_{:,i}=F^{H}\{\mathbf{Y}_{:,i}\}$\\
 
\hspace{1cm} $\widehat{k}_{j}=\underset{k}{\arg\max}\,\,\,\frac{\left|\left\langle D_{k},\widehat{\mathbf{X}}_{j,:}\right\rangle\right| }{||D_{k}||_{2}}$\\
 
\hspace{1cm}$\widehat{PD}^{j}=\max\left\{ \frac{\text{real}\left\langle D_{\widehat{k}_{j}},\widehat{\mathbf{X}}_{j,:}\right\rangle\ }{||D_{\widehat{k}_{j}}||_{2}^{2}},0\right\}$\\

\hspace{1cm}$\widehat{T}_{1}^{j},\widehat{T}_{2}^{j}=\mathbf{LUT}(\widehat{k}_{j})$
\label{alg1}
\end{algorithm}

Davies et al. \cite{davies2014compressed} suggested a method incorporating sparsity of the data in the dictionary domain (i.e. each pixel is represented by at most one dictionary item), referred to as the BLoch response recovery via Iterative Projection (BLIP) algorithm. This approach is based on the Projected Landweber Algorithm (PLA) which is an extension of the popular iterative hard thresholding method. 
BLIP (described here as \mbox{Algorithm 2}) consists of iterating between two main steps: A gradient step that enforces consistency with the measurements, and a projection that matches each row of $\mathbf{X}$ to a single dictionary atom.

 

BLIP and a few other works that are based on it \cite{wang2016magnetic} do not incorporate the temporal similarity across time points, which is a fundamental characteristic of the MRF sequence. In addition, there is a high degree of similarity across signatures in $\mathbf{D}$. As a result, the imaging contrasts matrix $\mathbf{X}$ is typically a low-rank matrix.

\begin{figure}
\begin{center}
{\hspace*{0cm} Singular values}\\
\begin{turn}{90}\parbox{7.5cm}{\hspace{0cm}\vspace{0cm} Log values}\end{turn}\hspace{1mm}\includegraphics[width=12cm,trim=2cm 1cm 1cm 1.8cm, clip=true]{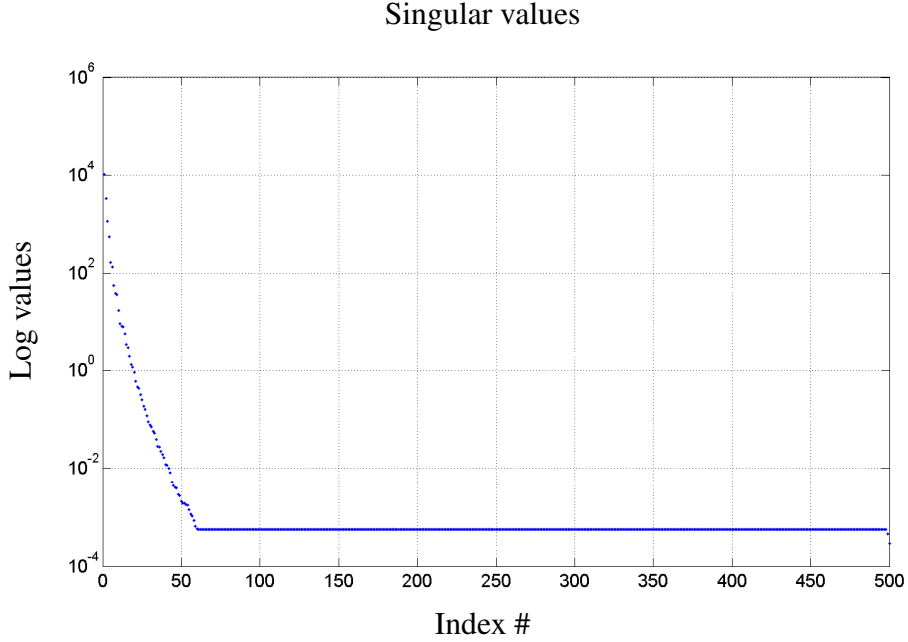}\\
{\hspace*{0cm} Index \#}\\
\caption{Singular values of FISP MRF sequence images, in descending order, presenting an effectively low rank matrix.}
\label{SV}
\end{center}
\end{figure}

Low-rank based modelling for dynamic MRI in general \cite{feng2013highly,chiew2015k} and MRF in particular \cite{mcgivney2014svd}-\cite{asslander2017low} has been proposed in the past.
To demonstrate the low-rank property of $\mathbf{X}$, we used T1, T2 and PD maps of size $128\times128$ (acquired using DESPOT \cite{deoni2005high}) as an input to a simulation of a FISP sequence \cite{jiang2015mr}, using $L=500$ TRs. In addition, we used random TR and FA values that have been used in previous publications in the field of MRF \cite{ma2013magnetic,jiang2015mr}. Note that the general assumption of $\mathbf{X}$ being a low-rank matrix holds as long as temporal similarity exist between time-frames in $\mathbf{X}$, and multiple voxels in the image belong to a single tissue, regardless of the specific acquisition parameters. Figure~\ref{SV} shows the singular values of $\mathbf{X}$. It can be seen that $\mathbf{X}$ is indeed low-rank, as most of the data is represented in the highest 15 singular values. 

This low-rank property of $\mathbf{X}$ can be exploited for improved reconstruction using the following optimization problem:
\begin{equation} 
\begin{aligned} 
& \underset{\mathbf{X,R}}{\text{minimize}} & & \frac{1}{2}\underset{i}{\Sigma}\left\Vert \mathbf{Y}_{:,i}-F_u\{\mathbf{X}_{:,i}\}\right\Vert _{2}^{2} \\ 
& \text{subject to} 
& & \textrm{rank}(\mathbf{X})\leq r \\
& & &\mathbf{X}=\mathbf{R}_1\mathbf{D}
\end{aligned} 
\label{eq5}
\end{equation}

\noindent where $\mathbf{R}_1$ is a matrix that matches each pixel ($\mathbf{X}_{j,:}$) with the dictionary signatures. In many previous implementations of low-rank for MRF, a matching of a single dictionary atom to a single pixel is enforced, which means that the rows of $\mathbf{R}_1$ are one-sparse vectors. The parameter $r$ is the rank of the matrix, and is usually defined as a fixed pre-chosen parameter. Typically $r$ is not known in advance and determining it upfront arise difficulty and may add error to the reconstruction scheme.

\begin{algorithm}[H]
\caption{BLIP}
\textbf{Input:}\\
A set of under-sampled k-space images: $\mathbf{Y}$\\ 
A pre simulated dictionary: $\mathbf{D}$\\ 
An appropriate look up table: $\mathbf{LUT}$\\
\textbf{Output:}\\
Magnetic parameter maps: $\widehat{T}_{1}, \widehat{T}_{2}, \widehat{PD}$\\
\textbf{Initialization:} $\mu, \widehat{\mathbf{X}}^0=\mathbf{0}$

\textbf{Iterate until convergence:}
\begin{itemize}
 \setlength{\itemindent}{-1em}
 
\item \hspace{-3mm} Gradient step for every $i$:

\hspace{-5mm} $\widehat{\mathbf{Z}}_{:,i}^{n+1}=\widehat{\mathbf{X}}_{:,i}^{n}-\mu F^{H}\{F_u\{\widehat{\mathbf{X}}_{:,i}^{n}\}-\mathbf{Y}_{:,i}\}$
\vspace{2mm}

\end{itemize}
\begin{itemize}
\setlength{\itemindent}{-1em}
\item \hspace{-3mm} Project onto the dictionary subspace for every $j$:

\hspace{-7mm} $ \widehat{k}_{j}=\underset{k}{\arg\max}\,\,\,\frac{\left|\left\langle D_{k},\widehat{\mathbf{Z}}_{j,:}\right\rangle\right| }{||D_{k}||_{2}}$

\hspace{-7mm} $ \widehat{PD}^{j}=\max\left\{ \frac{\text{real}\left\langle D_{\widehat{k}_{j}},\widehat{\mathbf{Z}}_{j,:}\right\rangle }{||D_{\widehat{k}_{j}}||_{2}^{2}},0\right\}$

\hspace{-7mm} $\widehat{\mathbf{X}}_{j,:}^{n+1}=\widehat{PD}^{j}\mathbf{D}_{\widehat{k}_{j}}$
\end{itemize}
\textbf{Restore maps for every $j$:} $\widehat{T}_{1}^{j},\widehat{T}_{2}^{j}=\mathbf{LUT}(\widehat{k}_{j}),\widehat{PD}^{j}$ 
\end{algorithm}

\begin{algorithm}[H]
\caption{Model Based Iterative Reconstruction MRF (MBIR-MRF)}
\textbf{Input:}\\
A set of under-sampled k-space images: $\mathbf{Y}$\\ 
A pre simulated dictionary: $\mathbf{D}$\\ 
An appropriate look up table: $\mathbf{LUT}$\\
\textbf{Output:}\\
Magnetic parameter maps: $\widehat{T}_{1}, \widehat{T}_{2}, \widehat{PD}$\\
\textbf{Initialization:} $\lambda, p<1, \eta_1, \eta_2, \mathbf{R}^0, \mat{Z}^0, \mat{Q}^0, \mat{W}^0=\mathbf{0},$\\ $\widehat{\mathbf{X}}_{:,i}^0=F^{H}\{\mathbf{Y}_{:,i}\}$ for every $i$

\textbf{Iterate until convergence:}

 \setlength{\itemindent}{-1em}

\begin{itemize} 
\item \hspace{-3mm} Find 1-sparse $\mathbf{R}$ as follows:\\

\hspace{5mm} $ \widehat{k}_{j}=\underset{k}{\arg\max}\,\,\,\frac{\left|\left\langle D_{k},\left(\widehat{\mathbf{X}}^{n}+\frac{1}{\eta_1}\mathbf{Q}^{n}\right)_{j,:}\right\rangle\right| }{||D_{k}||_{2}}$

\hspace{5mm} $ \mathbf{R}^{n+1}_{j,\widehat{k}_{j}}= \frac{\left\langle D_{\widehat{k}_{j}},\left(\widehat{\mathbf{X}}^{n}+\frac{1}{\eta_1}\mathbf{Q}^{n}\right)_{j,:}\right\rangle }{||D_{\widehat{k}_{j}}||_{2}^{2}}$
\end{itemize}

\setlength{\itemindent}{-1em}
\begin{itemize} 
\item \hspace{-3mm} Update $\mat{Z}^{n+1}$ by soft-thresholding the singular values:\\

\hspace{5mm} $[\mathbf{U},\mathbf{S},\mathbf{V}]=\mbox{svd}\left(\widehat{\mathbf{X}}^{n}+\frac{1}{\eta_2}\mathbf{W}^{n}\right)$ 

\vspace{2mm}
\hspace{5mm} Soft-threshold the non-zero singular values $\{\sigma\}_{j}$ of 

\hspace{5mm} $\mathbf{S}$ with parameter $\lambda\sigma_{j}^{p-1} $:

\hspace{5mm} $\sigma_{j}=\begin{cases}
\sigma_{j}-\lambda\sigma_{j}^{p-1} & \sigma_{j}>\lambda\sigma_{j}^{p-1}\\
0 & \mbox{otherwise}
\end{cases}$

\vspace{2mm}

\hspace{5mm} $\widehat{\mathbf{Z}}^{n+1}=\mathbf{USV}^{H}$
\end{itemize}

\setlength{\itemindent}{-1em}
\begin{itemize} 
\item \hspace{-3mm} Update $\mathbf{Q}$ and $\mathbf{W}$\\

\hspace{5mm} $\mathbf{Q}^{n+1} = \mathbf{Q}^{n}+\eta_1\left(\mathbf{X}^{n}-\mathbf{R}^{n+1}\mathbf{D}\right)$,
$\mathbf{W}^{n+1} = \mathbf{W}^{n}+\eta_2\left(\mathbf{X}^{n}-\mathbf{Z}^{n+1}\right)$
\vspace{2mm}
\end{itemize}
 \setlength{\itemindent}{-1em}

\begin{itemize} 
\item \hspace{-3mm} Update $\mat{X}^{n+1}$ by solving the following minimization problem:\\

\hspace{5mm} $\begin{aligned} 
& \underset{\mathbf{X}}{\text{minimize}} 
& &\frac{1}{2}\underset{i}{\Sigma}\left\Vert \mathbf{Y}_{:,i}-F_u\{\mathbf{X}_{:,i}\}\right\Vert _{2}^{2} +\eta_1\left\Vert \mathbf{Q}^{n+1}- \mathbf{X}+\mathbf{R}^{n+1}\mat{D}\right\Vert _F^2+\eta_2\left\Vert \mathbf{W}^{n+1}-\mathbf{X}+\mathbf{Z}^{n+1}\right\Vert _F^2\\ 
\end{aligned} $
\vspace{2mm}
\end{itemize}
\setlength{\itemindent}{-1em}
\textbf{Restore maps for every $j$:}\\ $\widehat{T}_{1}^{j},\widehat{T}_{2}^{j}=\mathbf{LUT}(\widehat{k}_{j}),\widehat{PD}^{j}=\max\left\{\mat{R}^{n+1}_{j,\widehat{k}_{j}},0\right\}$ 
\end{algorithm}

Zhao et al. \cite{zhao2015model} suggested an approximation for problem (\ref{eq5}), using an ADMM formulation \cite{boyd2011distributed} as follows:
\begin{equation} 
\begin{aligned} 
 \mathbf{X}^{n+1},\mathbf{R}_1^{n+1},\mathbf{Z}^{n+1} &=
& & \underset{\mathbf{X,R_{1},Z}}{\text{arg min}} \frac{1}{2}\underset{i}{\Sigma}\left\Vert \mathbf{Y}_{:,i}-F_u\{\mathbf{X}_{:,i}\}\right\Vert _{2}^{2}+\lambda\psi(\mathbf{Z})+\eta_1\left\Vert \mathbf{Q}^{n}- \mathbf{X}+\mathbf{R}_1\mathbf{D}\right\Vert _F^2\\
& & & +\eta_2\left\Vert \mathbf{W}^{n}-\mathbf{X}+\mathbf{Z}\right\Vert _F^2\\ 
 \mathbf{Q}^{n+1} &= 
& & \mathbf{Q}^{n}+\eta_1\left(\mathbf{X}^{n+1}-\mathbf{R}_1^{n+1}\mathbf{D}\right)\\
 \mathbf{W}^{n+1} &= 
& & \mathbf{W}^{n}+\eta_2\left(\mathbf{X}^{n+1}-\mathbf{Z}^{n+1}\right)
\end{aligned} 
\label{admm}
\end{equation}

\noindent where the low rank constraint is applied via the function $\psi(\mathbf{Z})$, defined as the $p$ norm ($p<1$) of the singular values of $\mathbf{Z}$ to the power of $p$. 
The matrices $\mathbf{Q}$ and $\mathbf{W}$ are the Lagrange multipliers.
The algorithm, coined Model Based Iterative Reconstruction MRF (MBIR-MRF) \cite{zhao2015model} is described in \mbox{Algorithm 3}.


\subsection{Proposed Method}

 The constraint presented in previous approaches \cite{zhao2015model} on $\mathbf{R}_1$ to have one sparse rows that contain the corresponding PD values for each row of $\mathbf{X}$, is justified by the assumption that only a single dictionary item should match an acquired signature. However, in practice, we found that superior results (in terms of spatial resolution and correspondence to ground truth) are obtained by relaxing this constraint, and allowing $\mathbf{X}$ to be comprised of multiple dictionary elements at each step of the optimization algorithm, where at the final stage each voxel is matched to a single tissue.  
 This allows for non-simulated signatures to be described by a linear combination of simulated ones. In addition, the relaxation enables formulating the problem as a convex problem, and saves the pattern recognition search time during reconstruction. The matching between $\mathbf{X}$ and the dictionary is done only at the final stage, after $\mathbf{X}$ is fully recovered by using a matched filter (MF), in order to extract the parameter maps. For brevity we write the constraint $\mathbf{X}=\mathbf{RD}$ as $\mathbf{X}\in \mathbb{D}$ where $\mathbb{D}=\{\mat{X}:\mathcal{N}(\mat{X})\supseteq \mathcal{N}(\mat{D})\}$
 , and we consider the next regularized form:
\begin{equation} 
\begin{aligned} 
& \underset{\mathbf{X \in \mathbb{D}}}{\text{minimize}} & & \frac{1}{2}\underset{i}{\Sigma}\left\Vert \mathbf{Y}_{:,i}-F_u\{\mathbf{X}_{:,i}\}\right\Vert _{2}^{2}+\lambda\textrm{rank}(\mathbf{X}) 
\end{aligned} 
\label{eq6}
\end{equation}
for some fixed regularization parameter $\lambda$.

Problem (\ref{eq6}) is not convex due to the rank constraint. We therefore relax this constraint by replacing the rank of $\mathbf{X}$ with the nuclear norm $\left\Vert \mathbf{X}\right\Vert _{*}$, defined as the sum of the singular values of $\mathbf{X}$ \cite{cai2010singular}. This results in the relaxed problem: 
\begin{equation} 
\begin{aligned} 
& \underset{\mathbf{X \in \mathbb{D}}}{\text{minimize}} & & \frac{1}{2}\underset{i}{\Sigma}\left\Vert \mathbf{Y}_{:,i}-F_u\{\mathbf{X}_{:,i}\}\right\Vert _{2}^{2} +\lambda\left\Vert \mathbf{X}\right\Vert _{*}.
\end{aligned} 
\label{eq7}
\end{equation}  
In order to solve (\ref{eq7}) we use the incremental subgradient proximal method \cite{sra2012optimization} as described in \mbox{Appendix A}. 

\begin{figure}
\begin{center}
{\hspace*{0cm} Reconstruction Error vs. Iterations}\\
\begin{turn}{90}\parbox{6cm}{\hspace{0.5cm}\vspace{0cm} Reconstruction error}\end{turn}\hspace{1mm}
\includegraphics[height=6cm,trim = 1.1cm 0.7cm 0.5cm 0.7cm, clip=true]{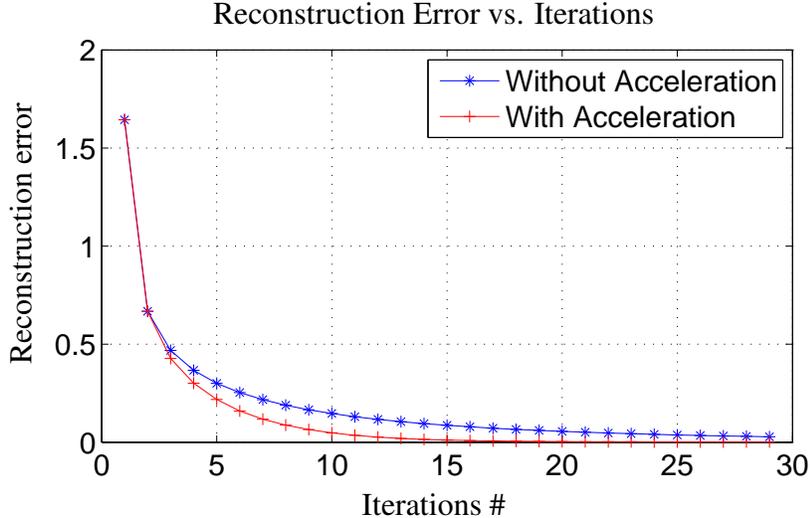}\\
{\hspace*{0cm} Iterations \#}\\
\caption{Comparison between the convergence of accelerated (plus sign) and standard (asterisk) FLOR.}
\label{Acc} 
\end{center}
\end{figure}

Due to the convex modelling of the problem, we also introduce an improvement that significantly reduces convergence time. The improvement uses the acceleration approach suggested by Nesterov \cite{nesterov1983method} for minimizing a smooth convex function, and its extension for non smooth composite functions of Beck and Teboulle  \cite{beck2009fast,palomar2010convex}. Our final algorithm is detailed in \mbox{Algorithm \ref{Acc}} and referred to as magnetic resonance Fingerprint with LOw Rank (FLOR), where the parameter $\lambda$ is chosen experimentally. Note that by setting $\lambda=0$, enforcing $\mathbf{R}$ to have one-sparse rows and eliminating the acceleration step, FLOR reduces to BLIP \cite{davies2014compressed}.

Figure~\ref{Acc} shows the reconstruction error of FLOR as the number of iterations varies with and without the acceleration step. Note that the CPU time of both algorithms is similar. 

\begin{algorithm}[H]
\caption{FLOR - MRF with LOw Rank}
\textbf{Input:}\\
A set of under-sampled k-space images: $\mathbf{Y}$\\
A pre simulated dictionary: $\mathbf{D}$\\ 
An appropriate look up table: $\mathbf{LUT}$\\
\textbf{Output:}\\
Magnetic parameter maps: $\widehat{T}_{1}, \widehat{T}_{2}, \widehat{PD}$\\
\textbf{Initialization:} $\mu$, $\lambda$, $t_{0}=1$, $\widehat{\mathbf{X}}^0=\mathbf{0}$,  $\mathbf{P}=\mathbf{D}^{\dagger}\mathbf{D}$

\textbf{Iterate until convergence:}

\begin{itemize}
\setlength{\itemindent}{-1em}
\item \hspace{-3mm} Gradient step for every $i$: 

\hspace{-5mm} $\widehat{\mathbf{Z}}_{:,i}^{n+1}=\widehat{\mathbf{X}}_{:,i}^{n}-\mu F^{H}\{F_u\{\widehat{\mathbf{X}}_{:,i}^{n}\}-\mathbf{Y}_{:,i}\}$
\vspace{2mm}
\setlength{\itemindent}{-1em}
\item \hspace{-3mm} Project onto the dictionary subspace and soft-threshold the singular values:
 
\hspace{-5mm} $[\mathbf{U},\mathbf{S},\mathbf{V}]=\mbox{svd}\left(\widehat{\mathbf{Z}}^{n+1}\mat{P}\right)$ 

\vspace{2mm}
\hspace{-6mm} Soft-threshold the non-zero singular values $\{\sigma\}_{j}$ of $\mathbf{S}$ with parameter $\lambda\mu $:

\hspace{-5mm} $\sigma_{j}=\begin{cases}
\sigma_{j}-\lambda\mu & \sigma_{j}>\lambda\mu\\
0 & \mbox{otherwise}
\end{cases}$

\vspace{2mm}

\hspace{-5mm} $\widehat{\mathbf{M}}^{n+1}=\mathbf{USV}^{H}$

\item \hspace{-3mm} Acceleration step:

\hspace{-5mm} $t_{n+1} = \frac{1+\sqrt{1+4t_{n}^{2}}}{2}$ 

\hspace{-5mm} $\widehat{\mathbf{X}}^{n+1}=\widehat{\mathbf{M}}^{n+1}+\frac{t_{n}-1}{t_{n+1}}\left(\widehat{\mathbf{M}}^{n+1}-\widehat{\mathbf{M}}^{n}\right)$ 
\end{itemize}
%
%
\textbf{Restore maps for every $j$:} 

\setlength{\itemindent}{-1em}
\hspace{5mm} $ \widehat{k}_{j}=\underset{k}{\arg\max}\,\,\,\frac{\left|\left\langle D_{k},\widehat{\mathbf{X}}_{j,:}\right\rangle\right| }{||D_{k}||_{2}}$
 
\hspace{5mm} $ \widehat{PD}^{j}=\max\left\{ \frac{\text{real}\left\langle D_{\widehat{k}_{j}},\widehat{\mathbf{X}}_{j,:}\right\rangle }{||D_{\widehat{k}_{j}}||_{2}^{2}},0\right\}$

\hspace{5mm}$\widehat{T}_{1}^{j},\widehat{T}_{2}^{j}=\mathbf{LUT}(\widehat{k}_{j})$ 
\end{algorithm}

\subsection{Possible extension}
 Conventional MRF algorithms use MF for the magnetic parameter extraction. MF introduces a quantization error since map values are continuous, as opposed to discrete dictionary values. A possible extension of FLOR is to add values to the dictionary by linear interpolation, in regions where a few candidates from the dictionary match a single signature from the data. We then select the dictionary signatures that exhibit a high correlation value (the ones above a certain threshold) and average their matching T1 and T2 values. 
This improvement expands the possible solutions to include ones that do not exist in the dictionary, and therefore exhibits improved accuracy compared to the conventional matching. The major benefit from this extension is reduced quantization errors that arise from conventional MF used in MRF. This extension, coined FLOR II, is examined in the first part of our experimental results in the next section.

\section{Experimental Results}
\label{sec:results}

\begin{figure}
\begin{centering}
\includegraphics[clip,width=0.5\columnwidth, trim=0cm 2cm 0cm 0cm]{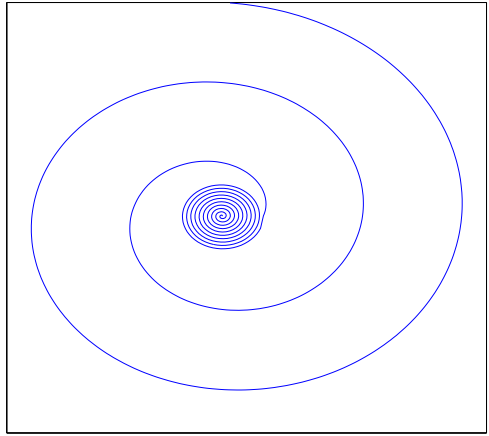}
\par\end{centering}
\caption{One of the spiral trajectories use for under-sampling a single image. Each time, the trajectories are rotated by 15 degrees.}
\label{spiral} 
\end{figure}

\begin{figure*}
\begin{centering}
\includegraphics[clip,width=1\textwidth,height=110mm, trim=5.8cm 2cm 3.5cm 0.5cm]{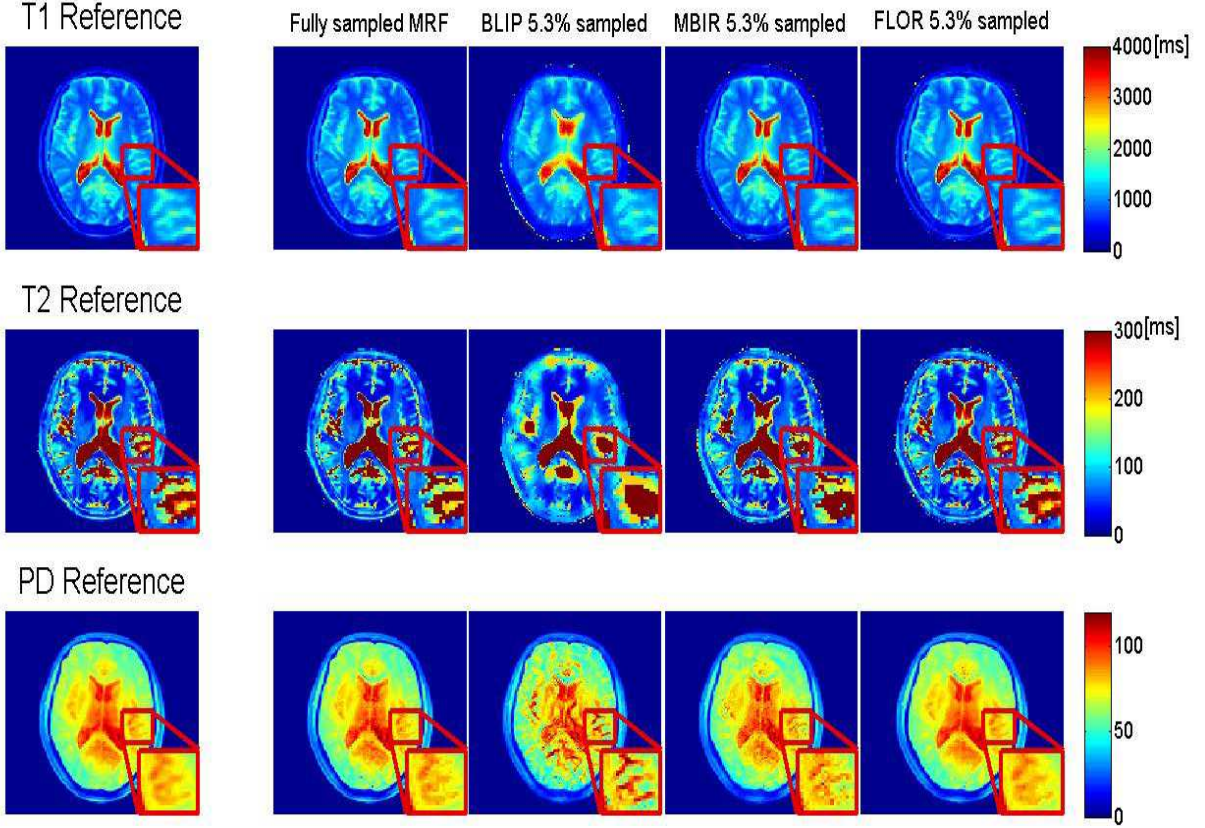}
\par\end{centering}
\caption{Reconstruction results of T1, T2 in milliseconds and PD in arbitrary units. Left: Reference maps, reconstruction using conventional MRF from 100\% of the noise free data, followed by BLIP, MBIR-MRF, and FLOR reconstruction with extension (as described in Section II.D) from 5\% of the noisy data.} 
\label{PM}
\end{figure*}
\begin{figure*}
\begin{centering}
\includegraphics[clip,width=0.9\textwidth,height=135mm, trim=4cm 1.7cm 2cm 0.9cm]{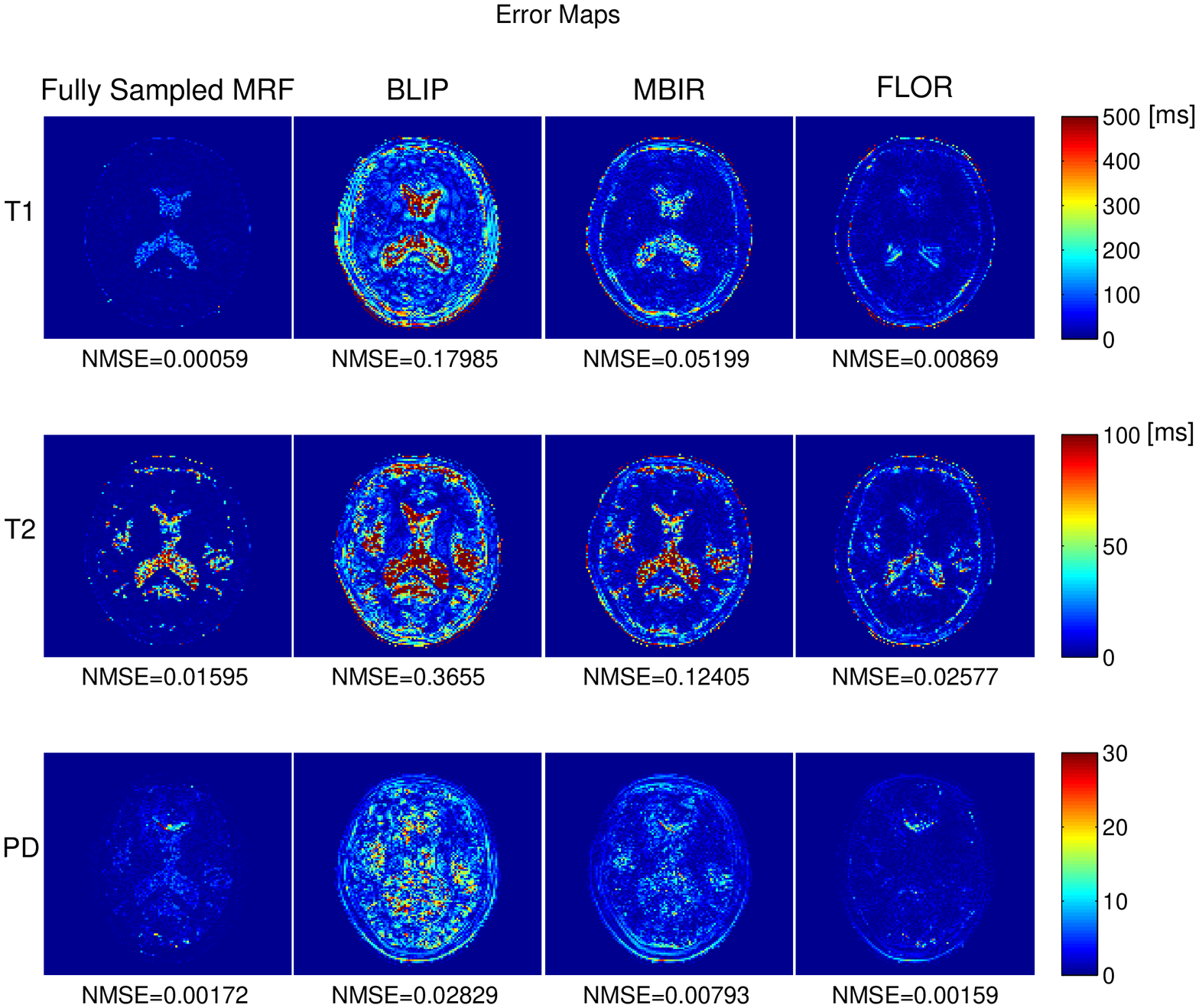}
\par\end{centering}
\caption{Error maps of the reconstruction of T1, T2 in milliseconds and PD in arbitrary units. Left: reconstruction using conventional MRF from 100\% of the noise-free data, followed by BLIP, MBIR-MRF, and FLOR reconstruction with extension (as described in Section II.D) from 5\% of the noisy data.}
\label{EM} 
\end{figure*}
This section describes two MRI experiments that were carried out using brain scans of a healthy subject. The first experiment is based on well known quantitative maps that were used, in a purely simulation environment to generate an MRF experiment with retrospectively sampled data. While this experiment is a simulation based on real quantitative maps, it allows accurate comparison of the results of the different algorithms using a well defined reference. 

In the second experiment, we used prospective sampled real MRF data that was used to generate the results in Ma et al. \cite{ma2013magnetic}. While this experiment lacks a gold-standard for accurate error evaluation, it allows comparison between different algorithms in a realistic multi-coil acquisition. To compare between different algorithms, for prospective sampling, where no ground truth is available, we examined the performance of the various algorithms as a function of the total number of excitations, where correspondence to values provided in literature for various brain tissues is used for validation. In both experiments, variable density spiral trajectories were used for sampling.

For quantitative error analysis, we calculated the normalized MSE (NMSE) between each quantitative map estimation and the reference map, defined as: 
\begin{equation} 
\begin{aligned}
\textrm{NMSE}_i= \frac{||\theta_{i}-\hat{\theta}_{i}||_F^2}{||\theta_{i}-\frac{1}{N}\underset{j}{\sum}\theta_{i}^{j}||_F^2}
\end{aligned}
\end{equation}
where $\theta_{i}$, $\hat{\theta}_i$ represent a reference map (such as T1,T2 or PD) and its corresponding reconstructed map (respectively), $N$ is the number of pixels in the map and $j$ is a spatial index. %


In the first experiment, forward and inverse non-uniform Fourier transforms were applied using SPURS, which is a fast approach published recently \cite{kiperwas2016spurs}. For the second experiment, we used the NUFFT package \cite{fessler2003nonuniform}, to adhere with the reconstruction results of the original MRF paper \cite{ma2013magnetic}.

\subsection{Experiment 1: Retrospective undersampling of real data}

The data for this experiment was acquired with a GE Signa 3T HDXT scanner. The
procedures involving human subjects described in this experiment were approved by the Institutional Review Board of Tel-Aviv
Sourasky Medical Center, Israel. We generated our reference data by acquisition of Fast Imaging Employing Steady-state Acquisition (FIESTA) and Spoiled Gradient Recalled Acquisition in Steady State (SPGR) images, at 4 different flip angles ($3^{\circ}$ ,$5^{\circ}$,$12^{\circ}$ and $20^{\circ}$), implementing the fast and well known DESPOT1 and DESPOT2 \cite{deoni2005high} algorithms, after improvements as described in Liberman et al. \cite{liberman2014t1}, to generate T1,T2 and PD quantitative maps, each of size $128 \times 128$ pixels. While it is well known that the gold standard method for T1 measurement is the inversion recovery spin echo with varying TIs and for T2 measurement is the spin echo sequences with varying TEs, in this experiment DESPOT was used as a reference thanks to its availability and its relatively fast acquisition time. The FISP pulse sequence has been applied for simulating acquisition of the reference. It was simulated with constant TE of 2ms, random TR values in the range of 11.5-14.5 ms, and a sinusoidal variation of FA (RF pulses) in the range of 0-70 degrees \cite{jiang2015mr}.

To simulate noisy undersampled MRF samples, we added complex Gaussian zero-mean noise to the k-space data to obtain an SNR of 67dB in the undersampled measurement domain. Data was then under-sampled to acquire only 876 k-space samples in each TR with spiral trajectories. In particular, we used 24 variable density spirals with inner region size of 20 and FOV of 24. In every time frame, each spiral is shifted by 15 degrees. Figure \ref{spiral} demonstrates the first spiral trajectory. We define the under-sampling ratio by the number of the acquired samples in the k-space domain divided by the number of pixels in the generated image. This leads to an undersampling ratio of \texttildelow5\% in this experiment. For comparison, the under-sampling ratio of the original MRF paper \cite{ma2013magnetic} is \texttildelow9\%, since for each single spiral 1450 data points were acquired.

We generated the dictionary using Bloch equations, simulating T1 values of [100:20:2000,\\2300:300:5000] ms and T2 values of [20:5:100,110:10:200,300:200:1900] ms. This range covers the relaxation time values that can be found in a healthy brain scan \cite{vymazal1999t1}. The tuning parameters were experimentally set as $\mu=1$ and $\lambda=5$, after $\lambda$ was tested in the range between 0 and 30.
Data was fed as an input to BLIP, MBIR-MRF and the improved FLOR algorithm (described as Algorithms 2,3 above and Algorithm 4 with the additional extension of interpolating the parameter maps). In addition, we performed reconstruction using 100\% of the data (without the addition of noise) via conventional MRF (\mbox{Algorithm 1}), for comparison purposes and to evaluate the error caused by the effect of discretized dictionary. All the iterative algorithms were run until the difference between consecutive iterations was below the same threshold.

The MATLAB code for reproducing the experiment provided in this section can be be found at: http://webee.techni- on.ac.il/Sites/People/YoninaEldar/software\_det18.php. In this code, spiral sampling trajectories design was based on Lee et al. \cite{lee2003fast}.


Figure~\ref{PM} shows the resulting maps for the recovery of T1, T2 and PD obtained with the various algorithms against the reference (left). The corresponding error maps of each method versus the reference are shown in Fig.~\ref{EM}. To allow detailed view of the reconstruction results for the reader, Fig.~\ref{PM} shows a zoomed region for each map. 

It can be seen that both FLOR and MBIR-MRF outperform BLIP reconstruction results, when using 5\% of sampled data by utilizing the low rank property. In addition, FLOR provides a lower error compared to MBIR-MRF. The details in the FLOR maps are comparable to those obtained by the original MRF algorithm using 100\% of the noise-free data. 
Due to the very low sampling ratio in our experiments (measured as the number of samples divided by the number of pixels in the image), conventional MRF using 5\% of the data did not provide valuable reconstruction results and is therefore omitted in this analysis. 

\begin{figure}
\begin{center} 
\includegraphics[width=1\columnwidth,trim= 1cm 2cm 0cm 1cm,clip=true]{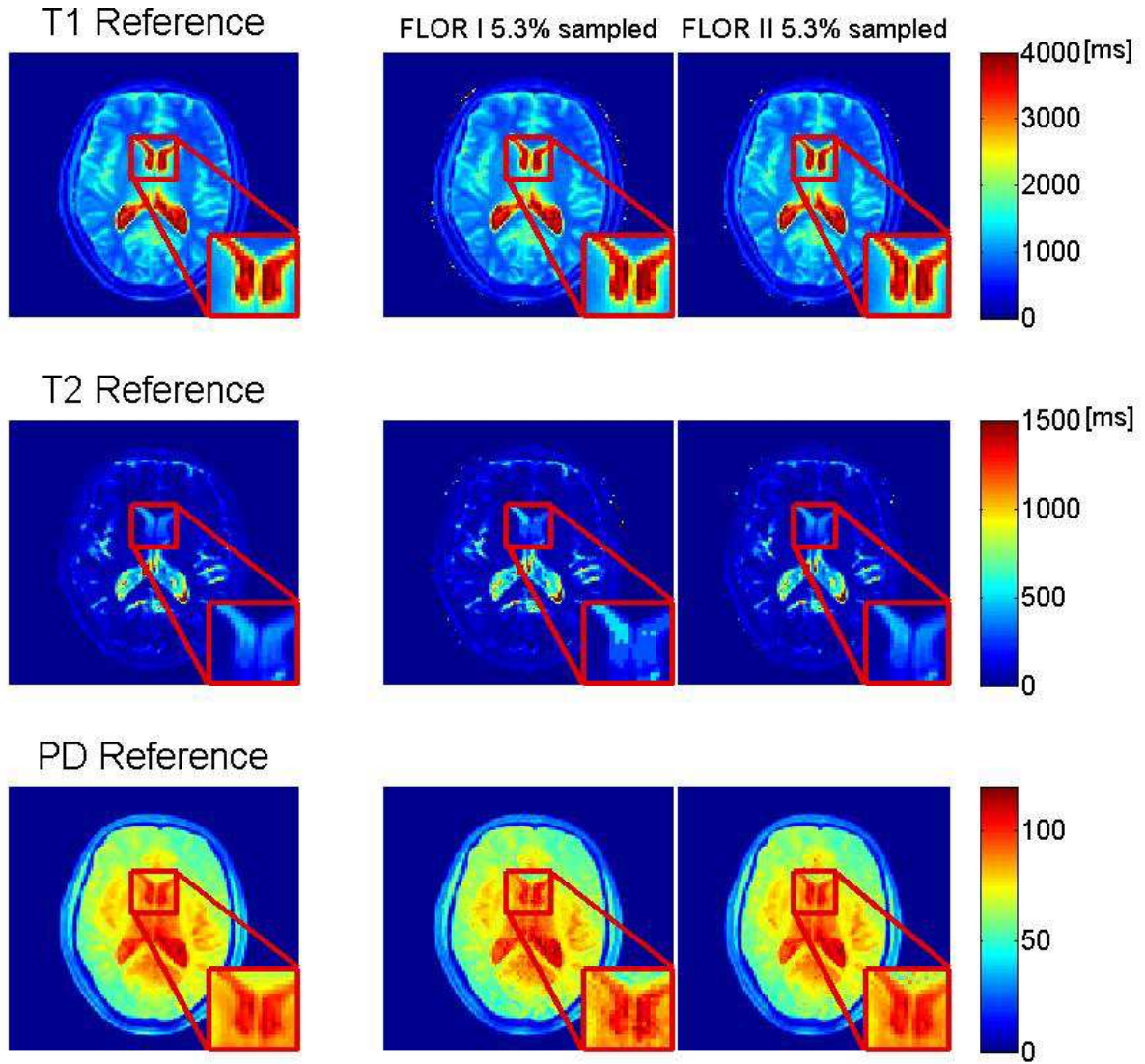}
\caption{Comparison of FLOR recovery with regular MF (FLOR I), and FLOR recovery with extension (FLOR II, as described in Section II.D) out of 5\% of the data. T1 and T2 maps are in milliseconds and PD is in arbitrary units. As can be seen in the magnified area, FLOR II demonstrates a smoother solution (with less quantization errors) which is more similar to the original reference maps.}
\label{FLOR2a}
\vspace{2mm}
\end{center}
\end{figure}
\begin{figure}
\begin{center}
{\hspace*{-1.5cm} Error maps}\\
\includegraphics[width=0.6\columnwidth,,height=135mm,trim=1cm 1.5cm 0cm 2cm,clip=true]{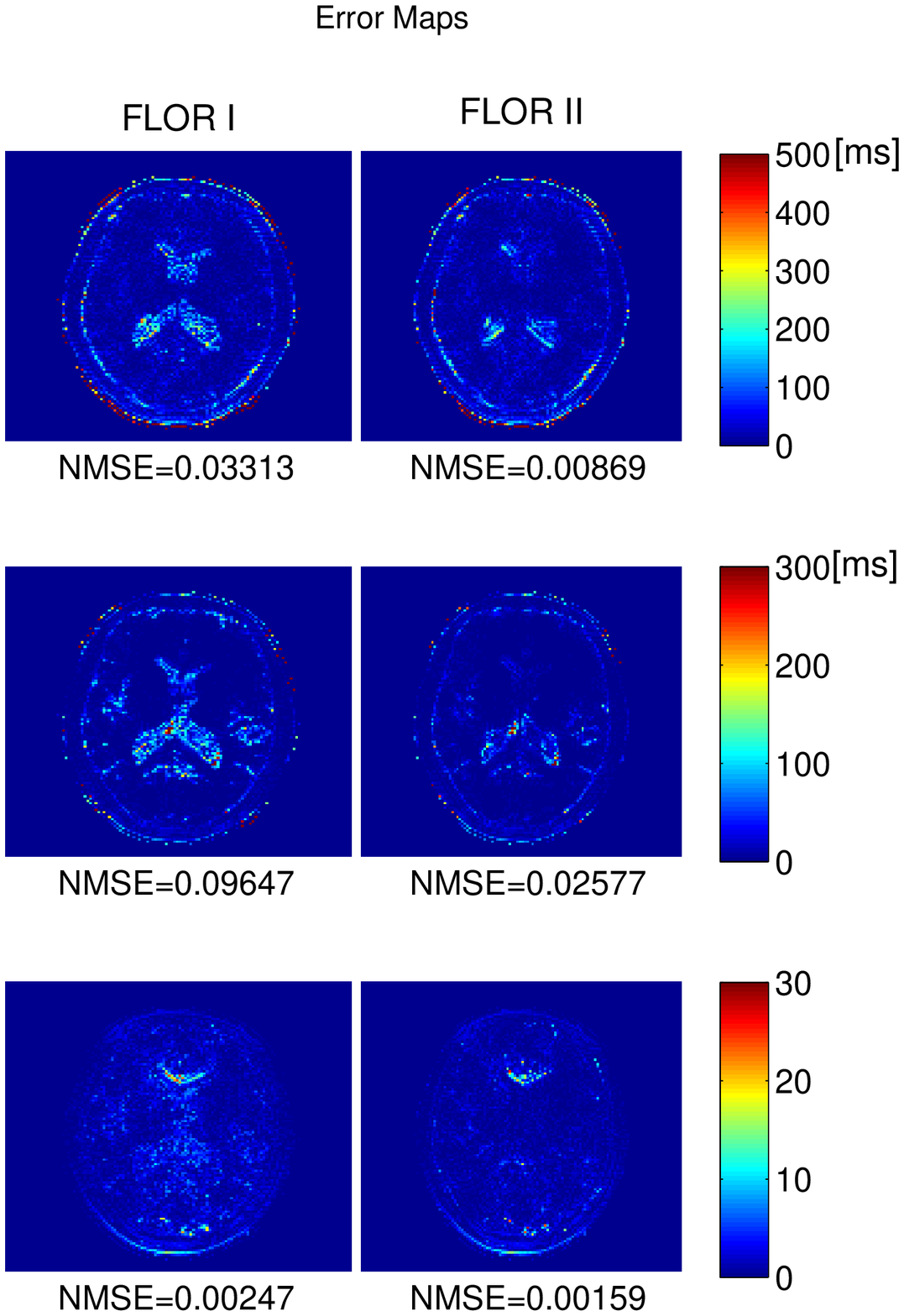}
\caption{Error maps comparison, FLOR I vs. FLOR II. The T1 and T2 error maps are in milliseconds and PD is in arbitrary units. It can be seen that FLOR II outperforms FLOR I.}
\label{FLOR2b} 
\vspace{2mm}
\end{center}
\end{figure}

We next implemented the MF improvement described in Section II.D. The results are shown in Fig.~\ref{FLOR2a}, with corresponding error maps in Fig.~\ref{FLOR2b}.  These figures compare the recovery maps of FLOR without (FLOR I) and with (FLOR II) the proposed improvement. 
It can be seen that FLOR II improves the results of FLOR I and produces a smoother solution which better fits the reference maps.

\subsection{Experiment 2: In vivo prospective sampling experiment}
The experiment in this section was carried out using the data of the original MRF paper \cite{ma2013magnetic}, 
which consisted of 48 spiral trajectories shifted by 7.5 degrees, where 1450 samples were acquired in each trajectory, leading to an underampling ratio of 9\%. The data was acquired on a 1.5-T whole body scanner (Espree, SIEMENS Healthcare) using a 32-channel head receiver coil. 

\begin{figure*}

{\hspace*{-5mm} T1 reference
\hspace{8mm} T2 reference \hspace{10mm} PD reference \hspace{12mm} df reference}\\
\hspace{-5mm}
\includegraphics[height=100pt,trim=1cm 0cm 1.2cm 0cm, clip=true]{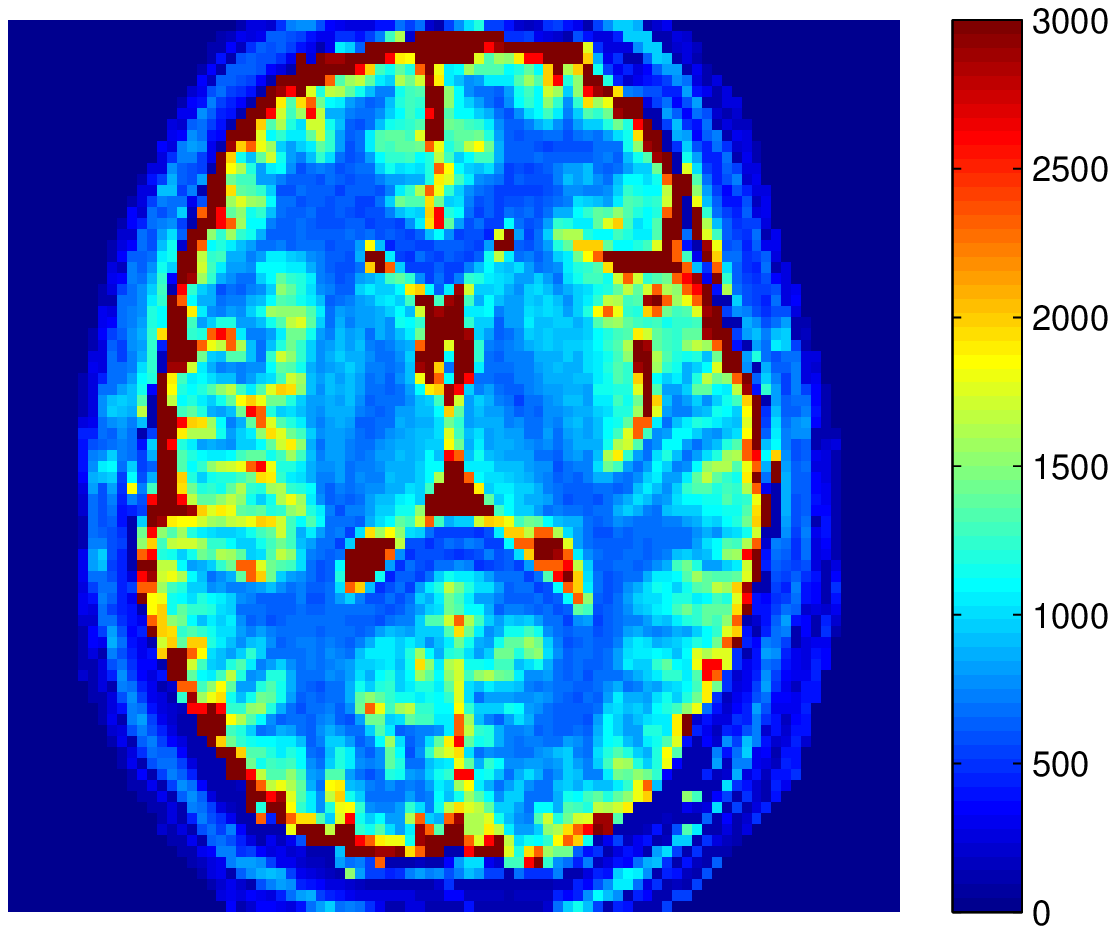}\hspace{-1mm}
\includegraphics[height=100pt,trim=1cm 0cm 1.2cm 0cm, clip=true]{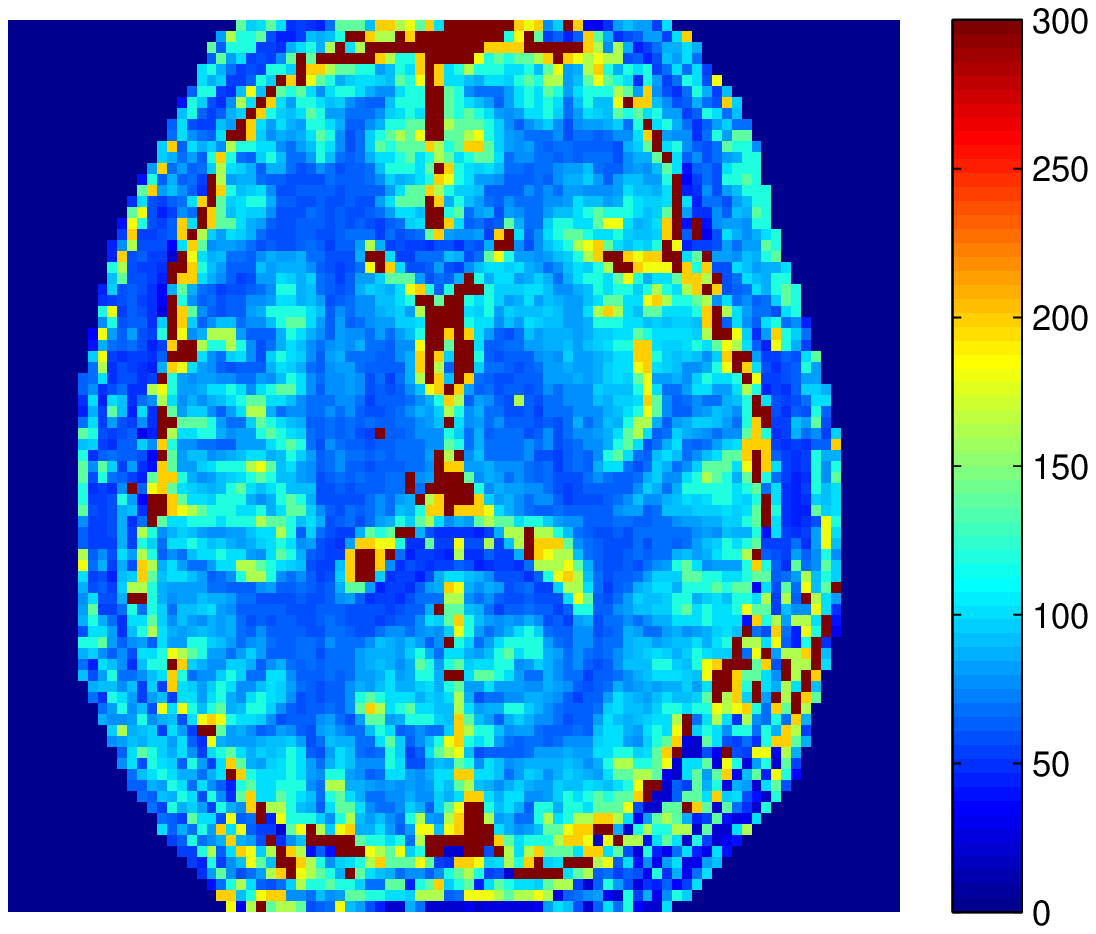}\hspace{-2mm}
\includegraphics[height=100pt,trim=1cm 0cm 1.2cm 0cm, clip=true]{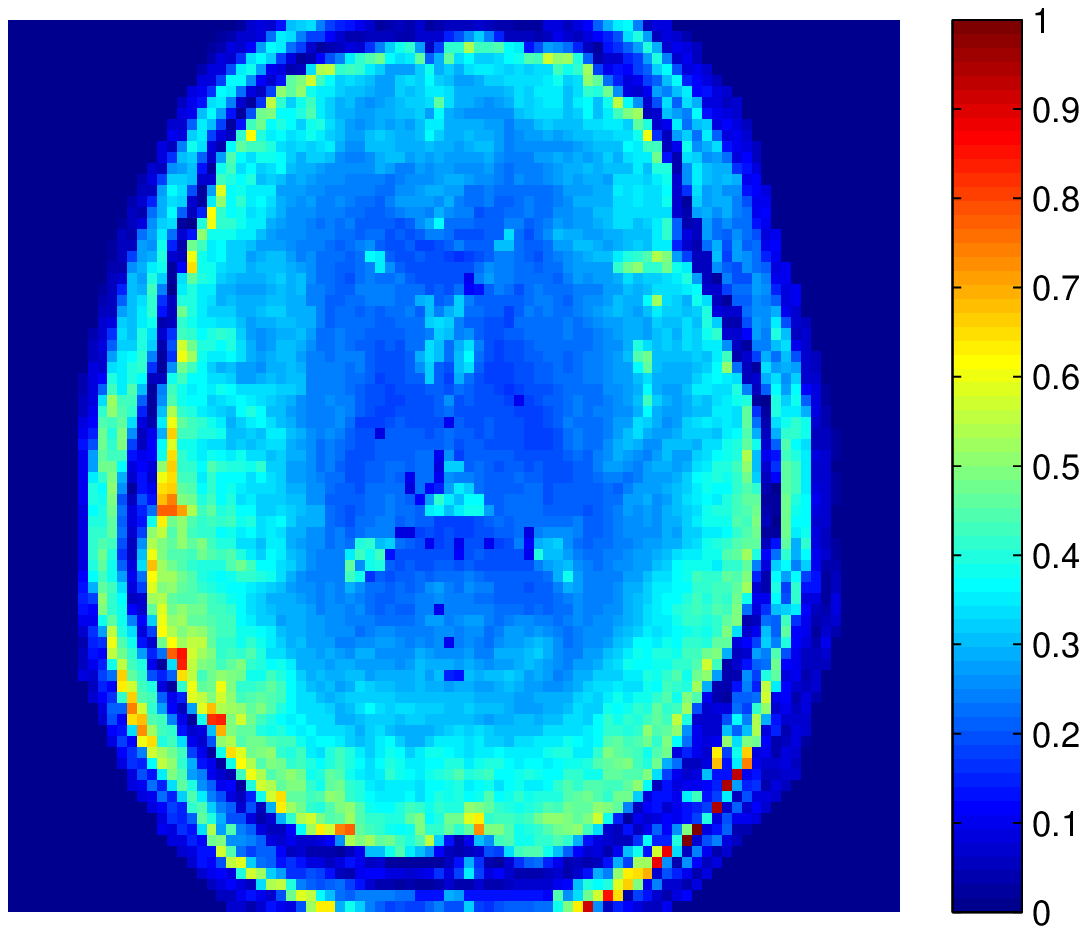}\hspace{-3mm}
\includegraphics[height=100pt,trim=1.5cm 0cm 1.2cm 0cm, clip=true]{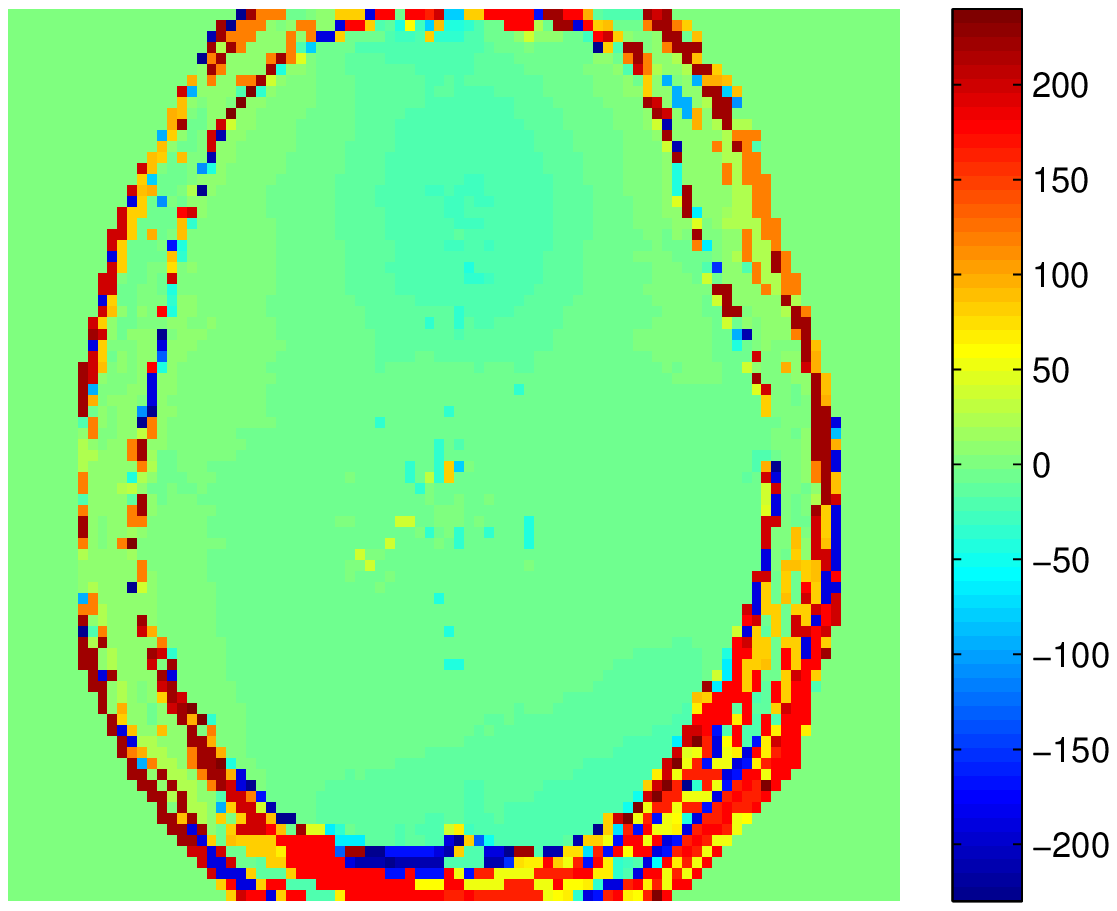}

  \caption{The maps obtained when applying the original MRF algorithm on 1000 TRs. T1 and T2 color scales are in milliseconds, PD in normalized color scale and df color scale in Hz. }
        \label{fig2_reference_MRF}
\end{figure*}

Due to the lack of gold standard maps for this data, we are unable to provide quantitative error results (e.g. NMSE). Therefore, in this experiment we compare between the various algorithms by examining reconstruction results using 400 TRs (representing 40\% of scanning time), 
to quantitative values of brain tissues from the literature. Since the results obtained in the original MRF experiment (using 1000 TRs) mostly correspond to quantitative values from the literature, the maps generated using 1000 TRs using the original MRF algorithm are provided in Fig.~\ref{fig2_reference_MRF}, for reference.

\begin{figure*}
{\hspace*{-5mm} BLIP with 32 coils
\hspace{10mm} MBIR MRF with 32 coils \hspace{12mm} FLOR with 32 coils}\\
\hspace{-10mm}
\begin{turn}{90}\parbox{5cm}{\hspace{0mm} T1 Reconstruction}\end{turn}\hspace{0mm}
\includegraphics[width=150pt,height=130pt,trim=2cm 0cm 1cm
0.5cm, clip=true]{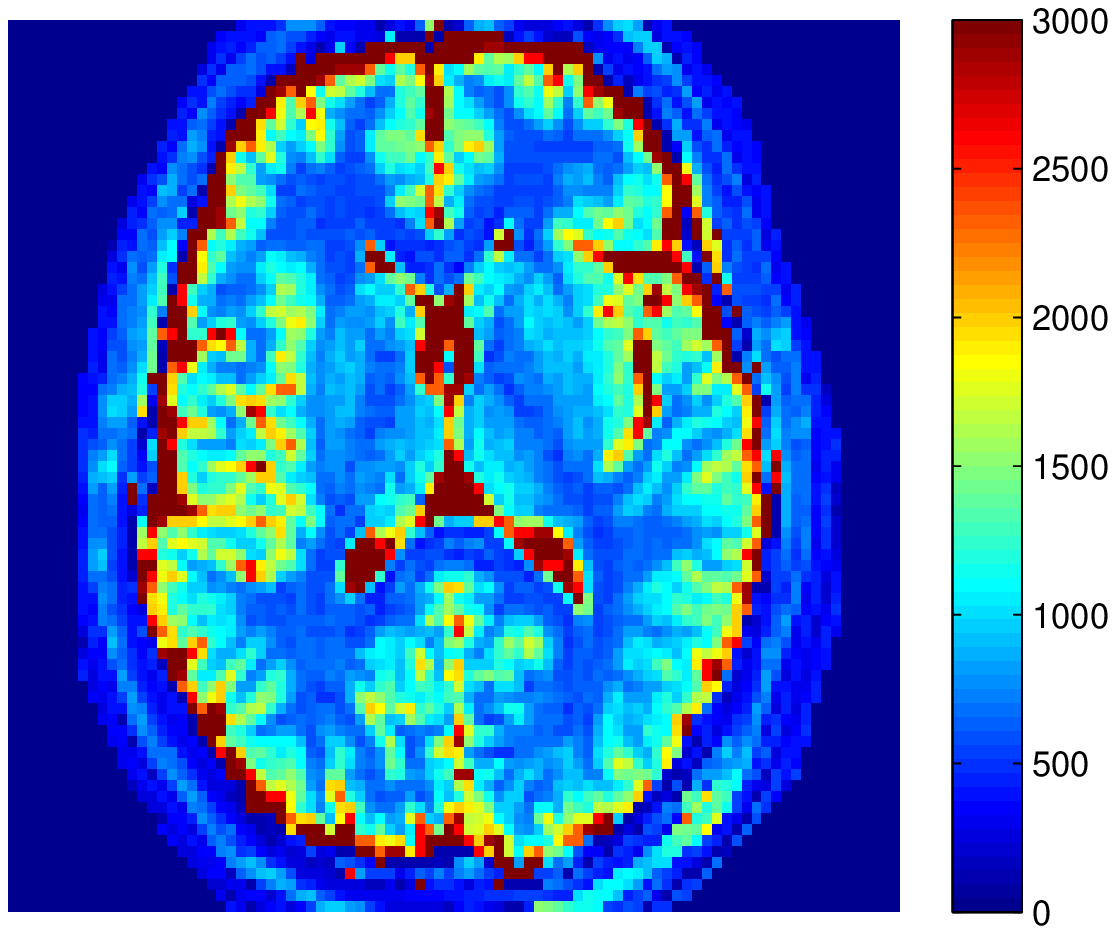}\hspace{0mm}
\includegraphics[width=150pt,height=130pt,trim=2cm 0cm 1cm
0.5cm, clip=true]{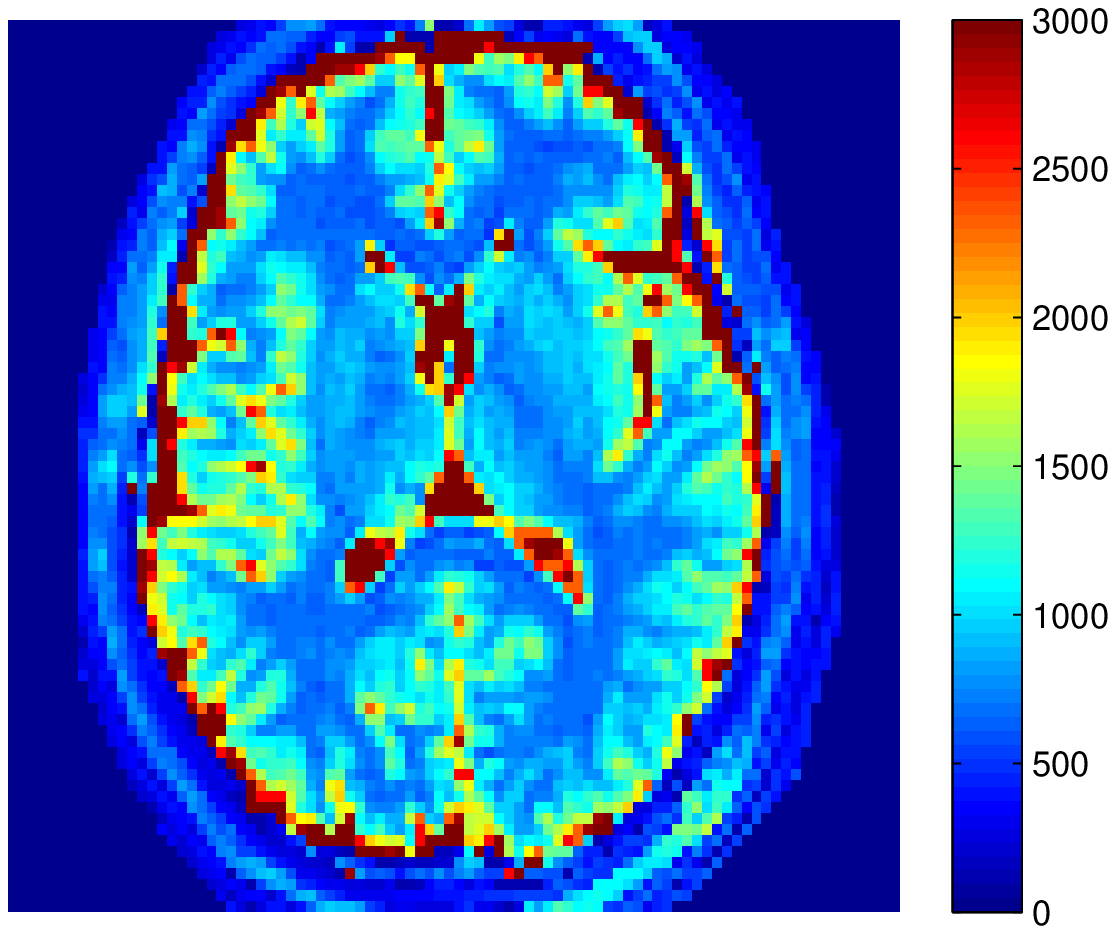}\hspace{1mm}
\includegraphics[width=150pt,height=130pt,trim=2cm 0cm 1cm
0.5cm, clip=true]{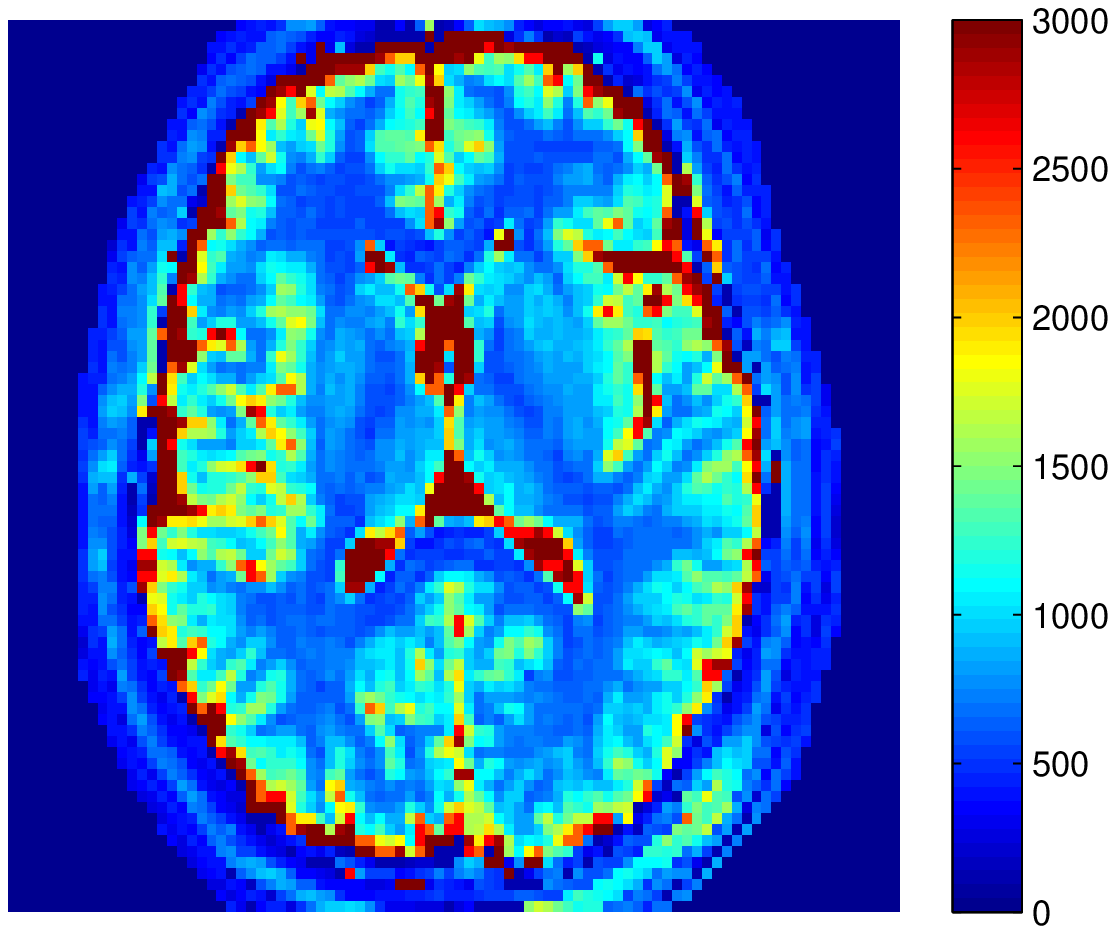}\vspace{0mm}\\
\hspace{-10mm}\begin{turn}{90}\parbox{5cm}{\hspace{0mm} T2 Reconstruction}\end{turn}\hspace{0mm}
\includegraphics[width=150pt,height=130pt,trim=2cm 0cm 1cm
0.5cm, clip=true]{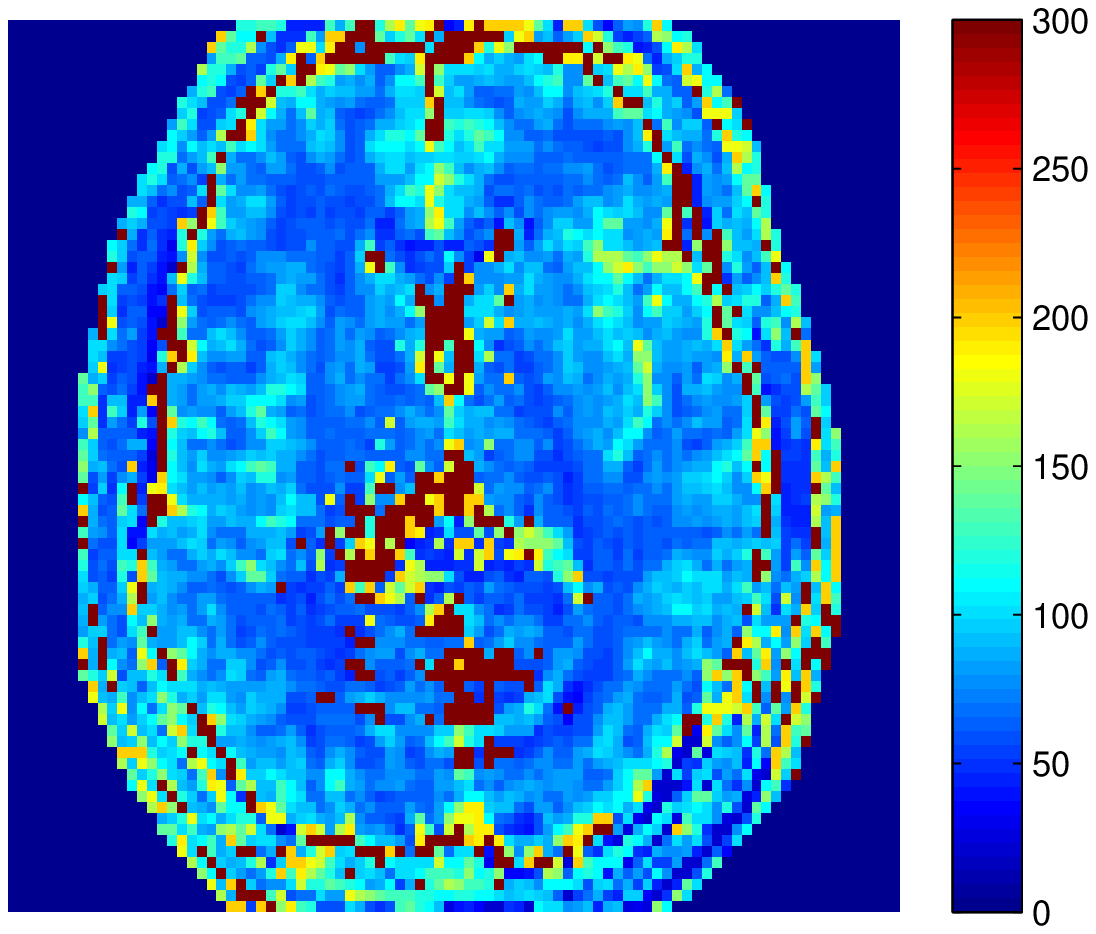}\hspace{0mm}
\includegraphics[width=150pt,height=130pt,trim=2cm 0cm 1cm
0.5cm, clip=true]{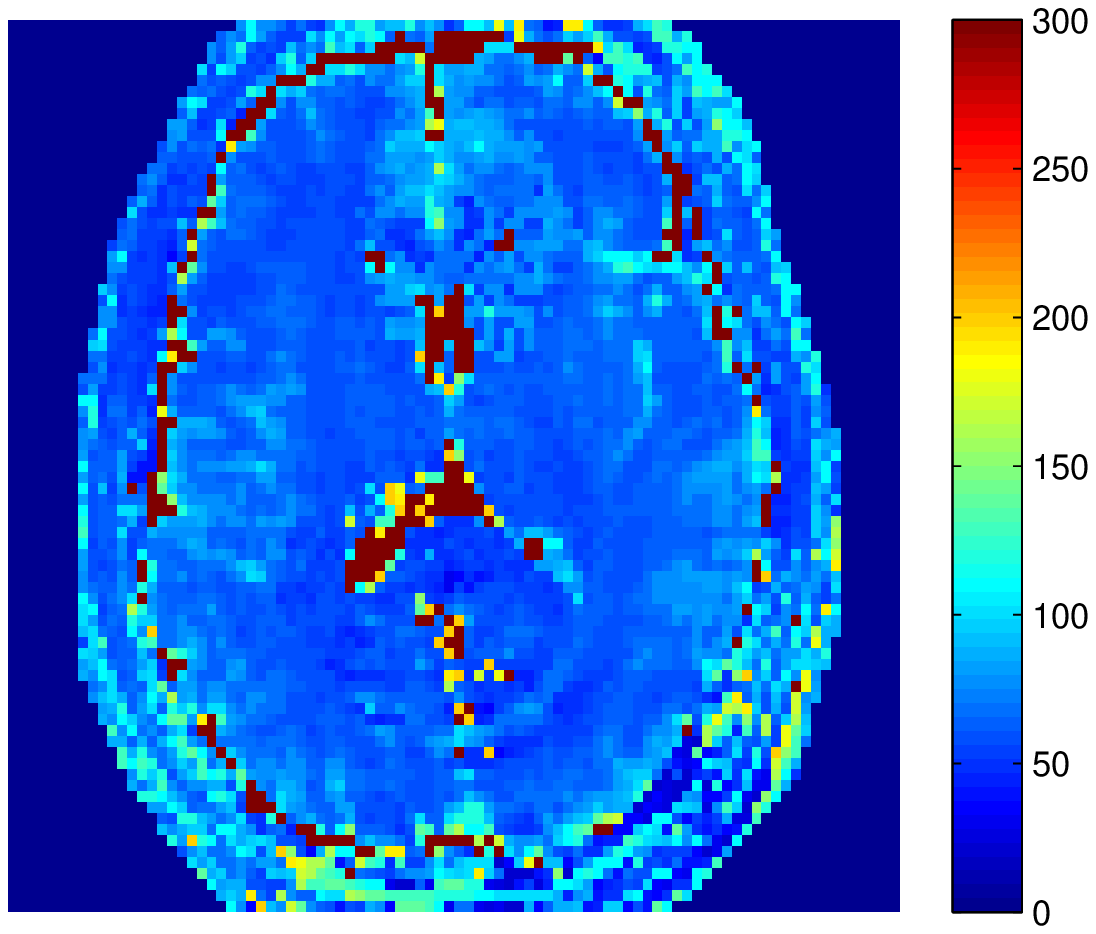}\hspace{1mm}
\includegraphics[width=150pt,height=130pt,trim=2cm 0cm 1cm
0.5cm, clip=true]{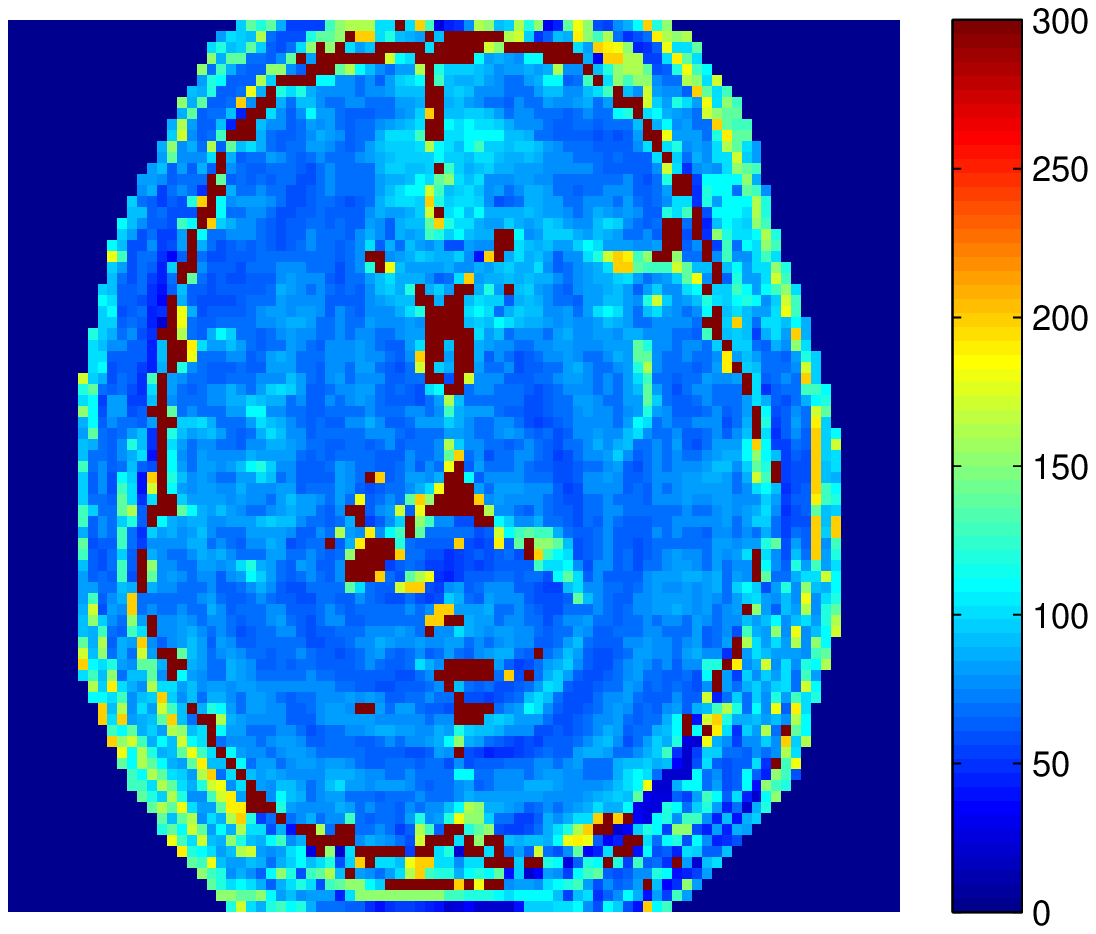}\vspace{0mm}\\
\hspace{-10mm}\begin{turn}{90}\parbox{5cm}{\hspace{0mm} PD Reconstruction}\end{turn}\hspace{0mm}
\includegraphics[width=150pt,height=130pt,trim=2cm 0cm 1cm
0.5cm, clip=true]{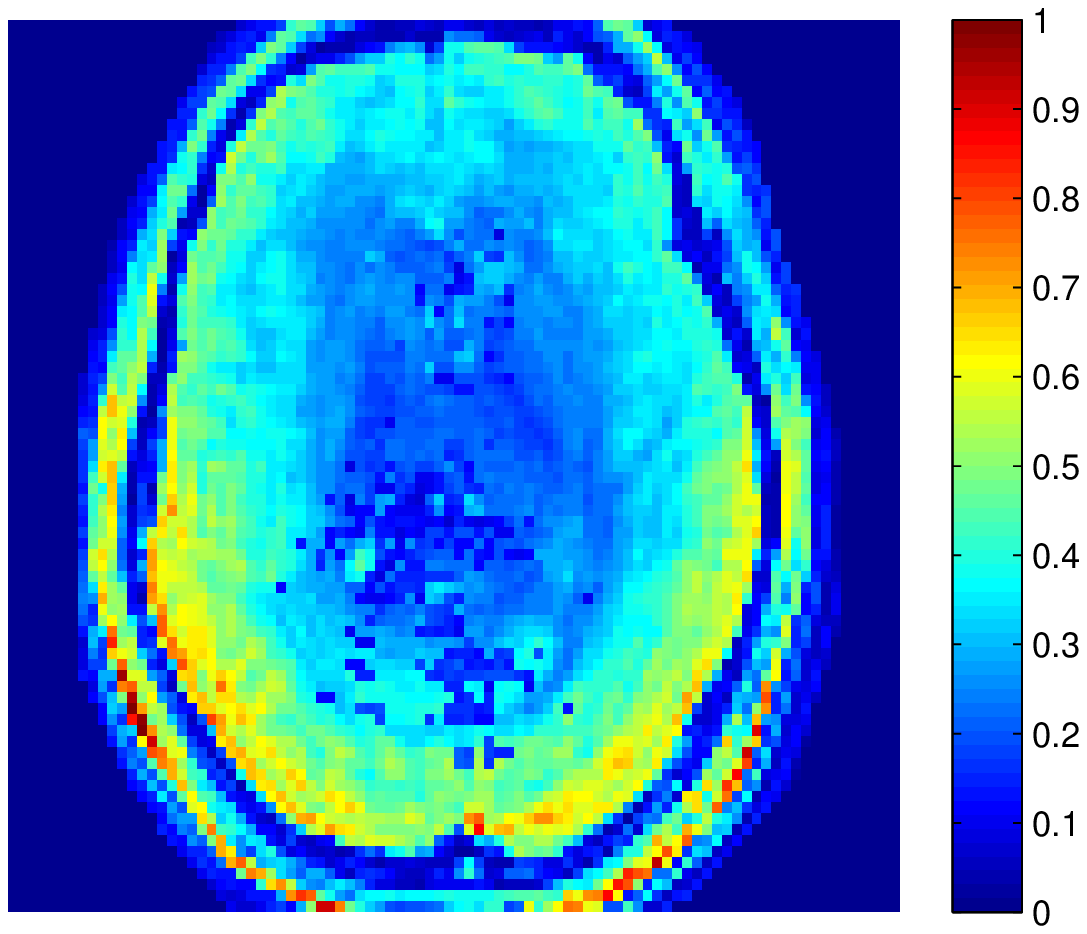}\hspace{0mm}
\includegraphics[width=150pt,height=130pt,trim=2cm 0cm 1cm
0.5cm, clip=true]{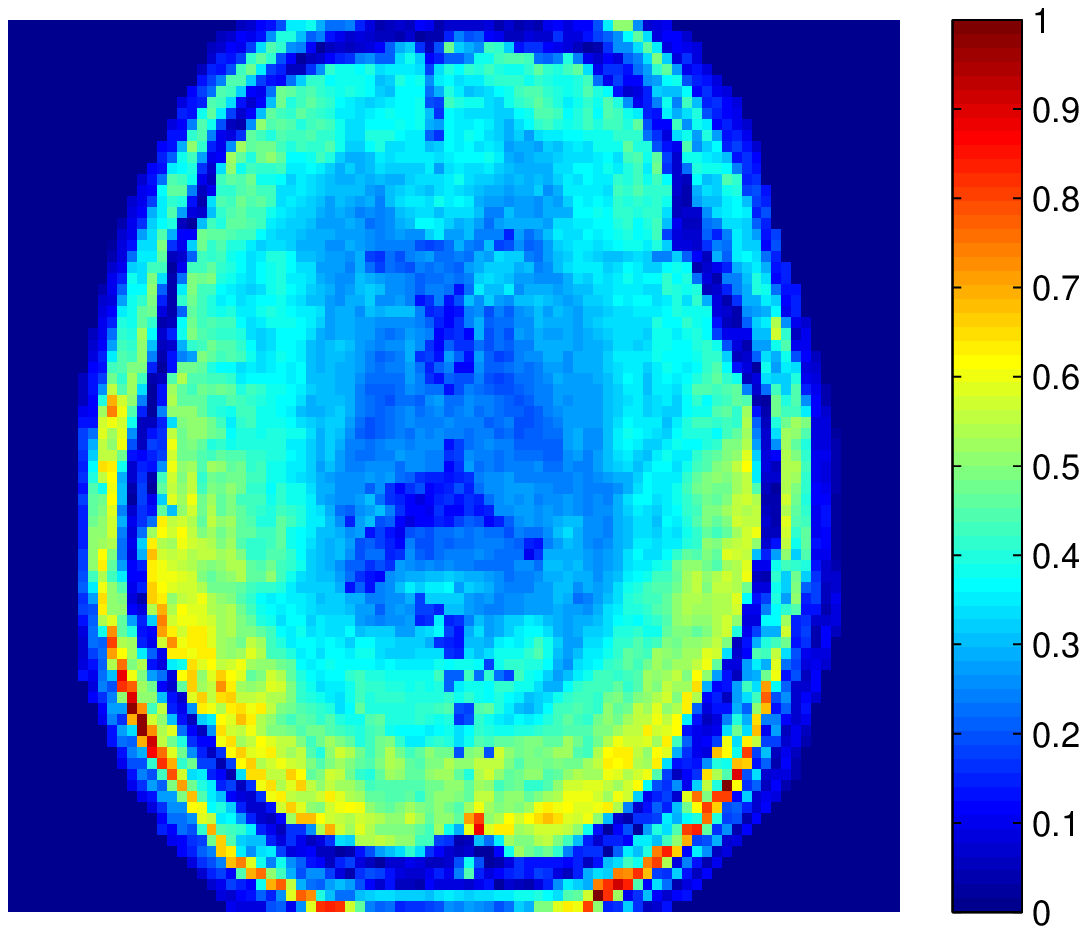}\hspace{1mm}
\includegraphics[width=150pt,height=130pt,trim=2cm 0cm 1cm
0.5cm, clip=true]{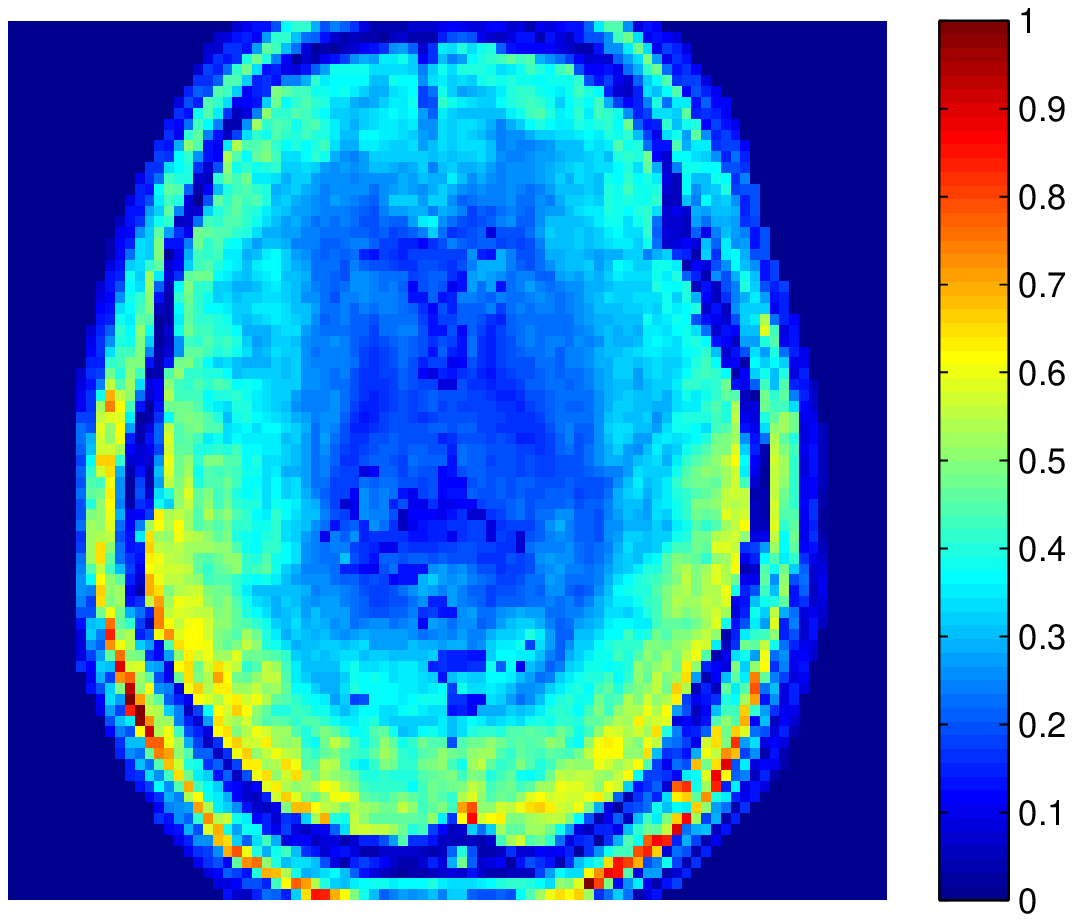}\vspace{0mm}\\
\hspace{-10mm}\begin{turn}{90}\parbox{5cm}{\hspace{0mm} df Reconstruction}\end{turn}\hspace{0mm}
\includegraphics[width=150pt,height=130pt,trim=2cm 0cm 1cm
0.5cm, clip=true]{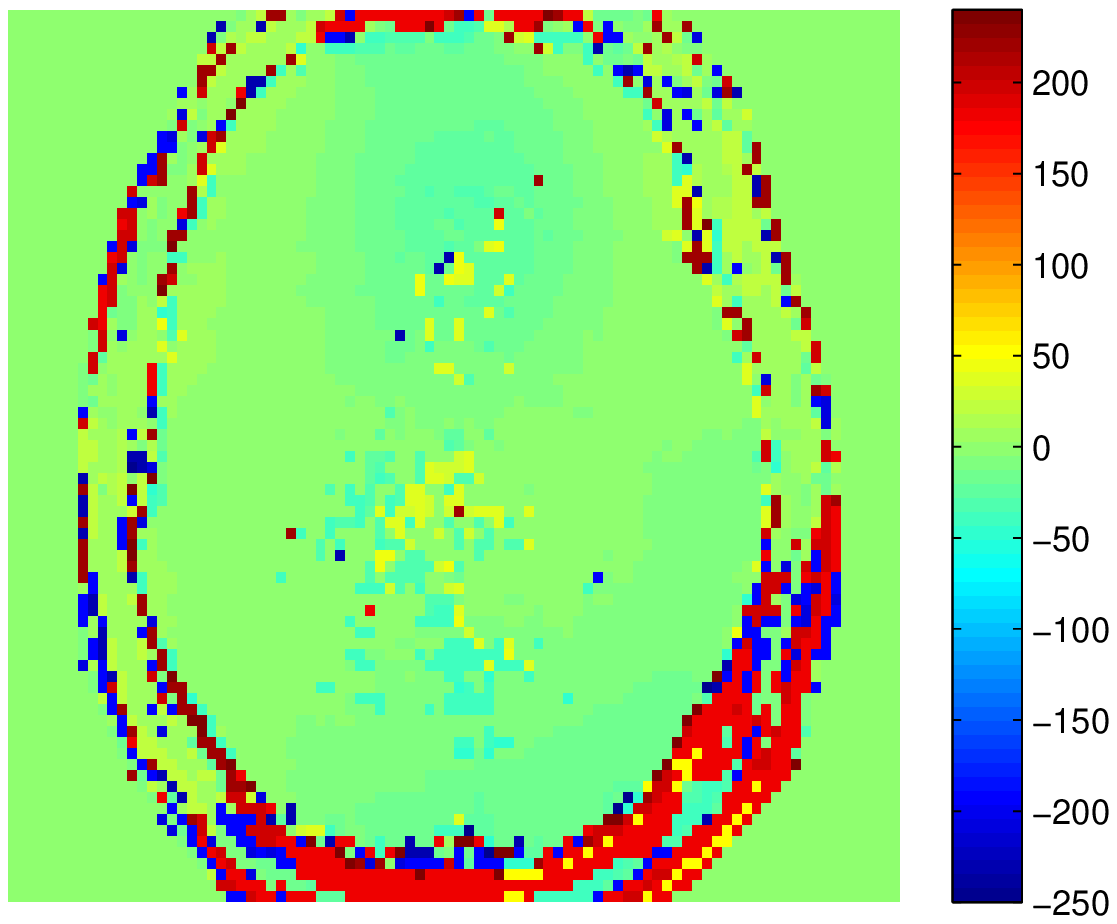}\hspace{0mm}
\includegraphics[width=150pt,height=130pt,trim=2cm 0cm 1cm
0.5cm, clip=true]{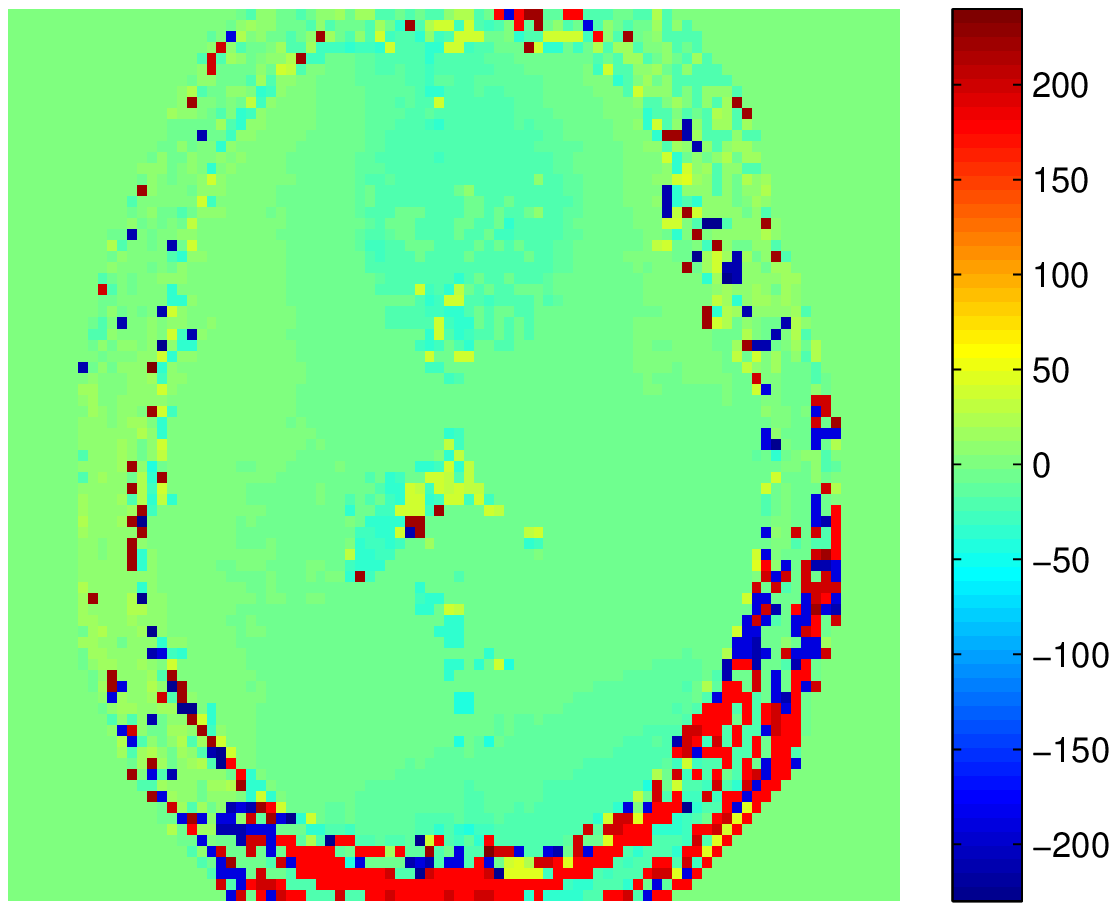}\hspace{1mm}
\includegraphics[width=150pt,height=130pt,trim=2cm 0cm 1cm
0.5cm, clip=true]{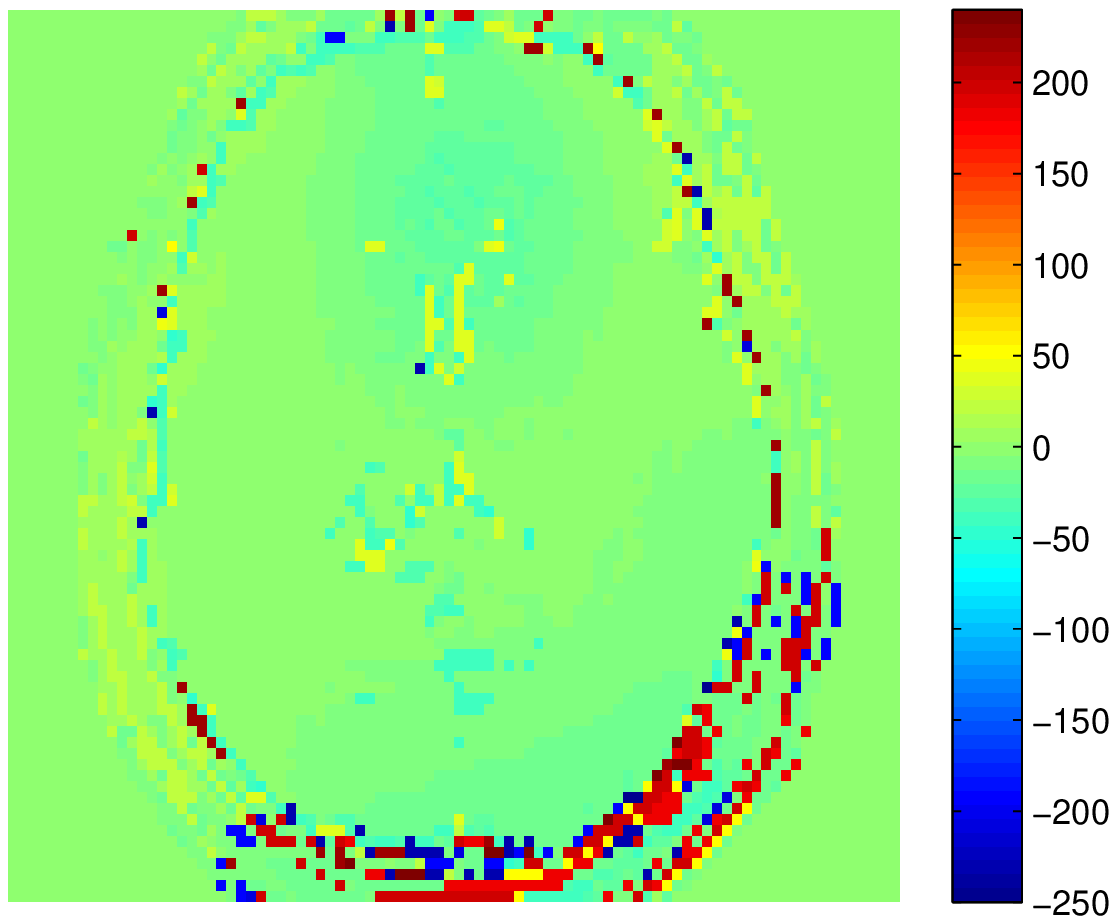}\\


  \caption{Comparison between BLIP, MBIR-MRF and FLOR reconstructions with data acquired out of 32 coils and 400 TRs. T1 and T2 color scales are in milliseconds, PD in normalized color scale and df color scale in Hz}
        \label{real_data}
\end{figure*}

\begin{figure*}
{\hspace{0mm} T2 Reconstruction}\\
\vspace{0.3cm}
{\hspace*{-10mm} BLIP with 32 coils
\hspace{10mm} MBIR MRF with 32 coils \hspace{10mm} FLOR with 32 coils}\\
\hspace{-10mm}
\includegraphics[width=150pt,height=130pt,trim=2cm 1cm 1cm
0.6cm, clip=true]{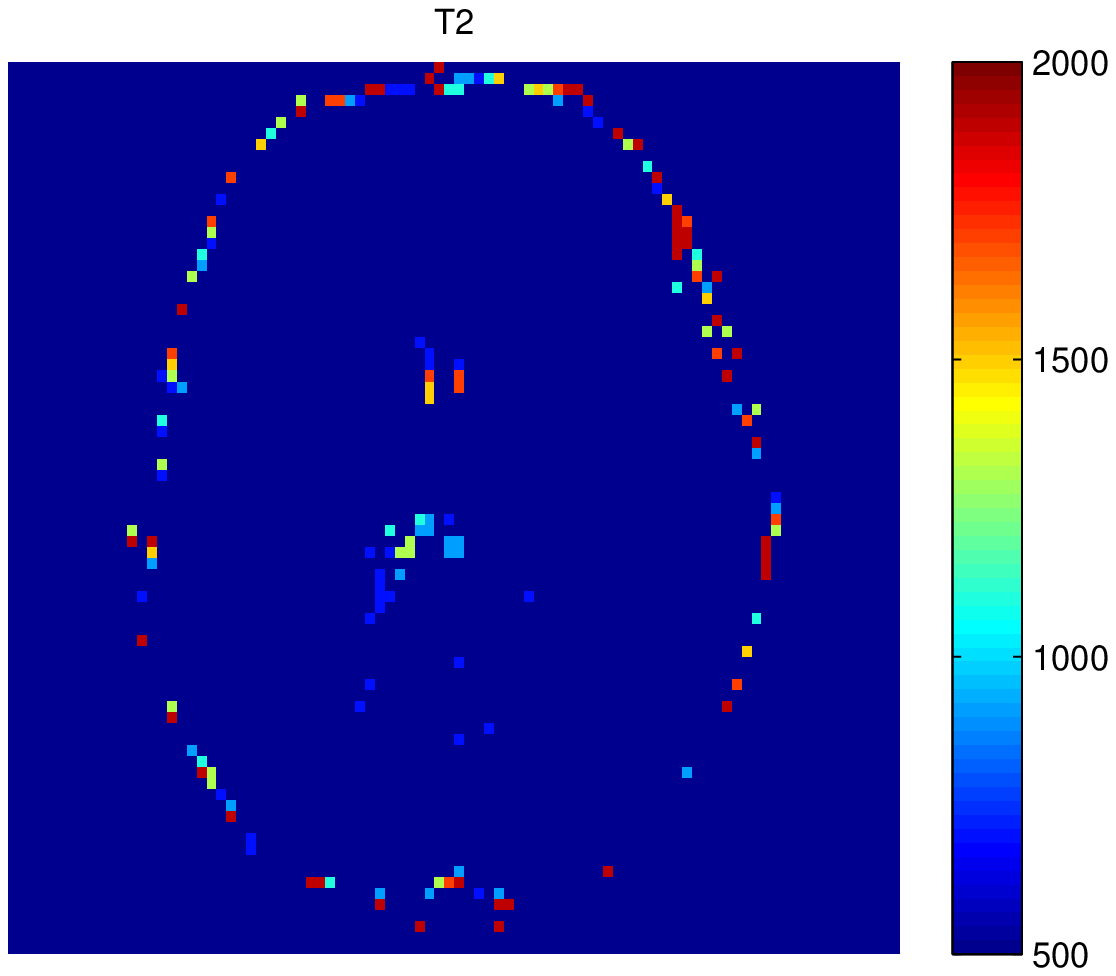}\hspace{0mm}
\includegraphics[width=150pt,height=130pt,trim=2cm 1cm 1cm
0.6cm, clip=true]{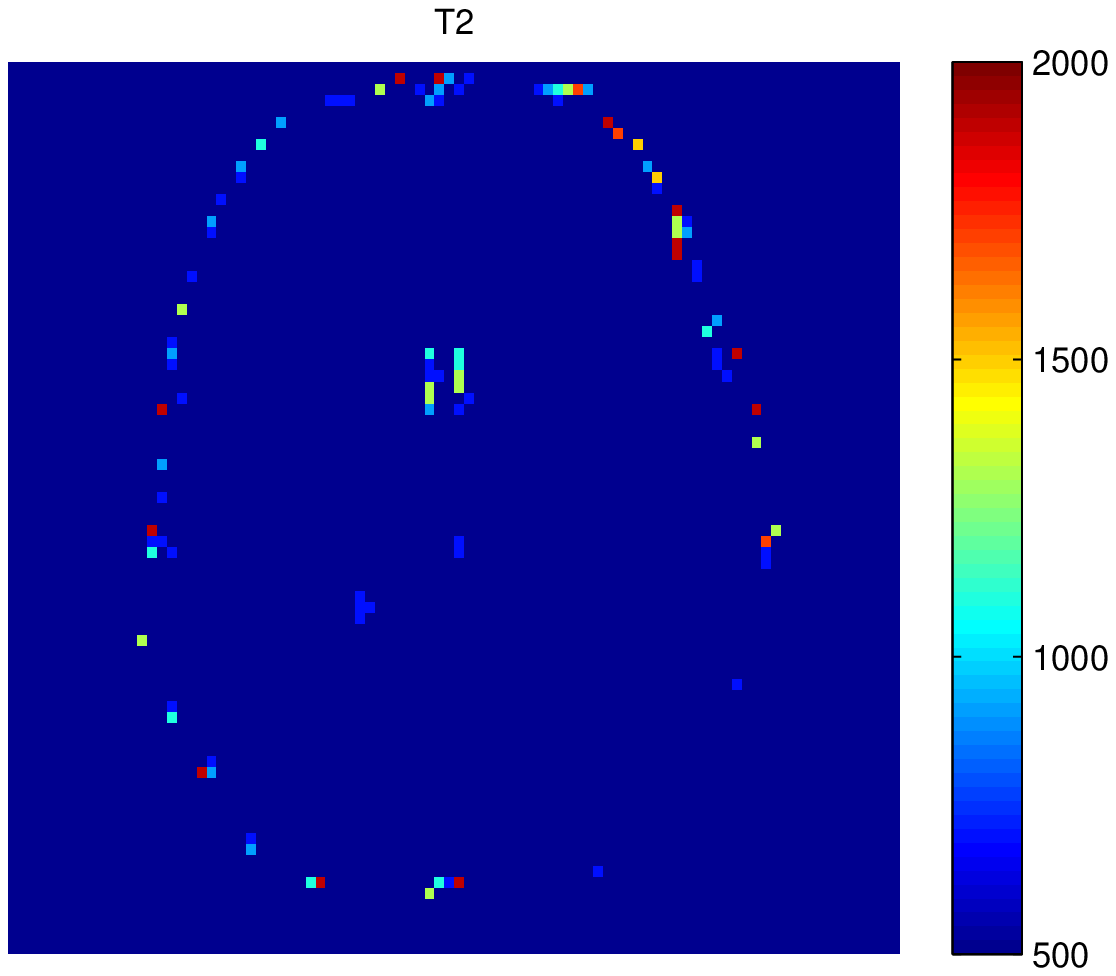}\hspace{1mm}
\includegraphics[width=150pt,height=130pt,trim=2cm 1cm 1cm
0.6cm, clip=true]{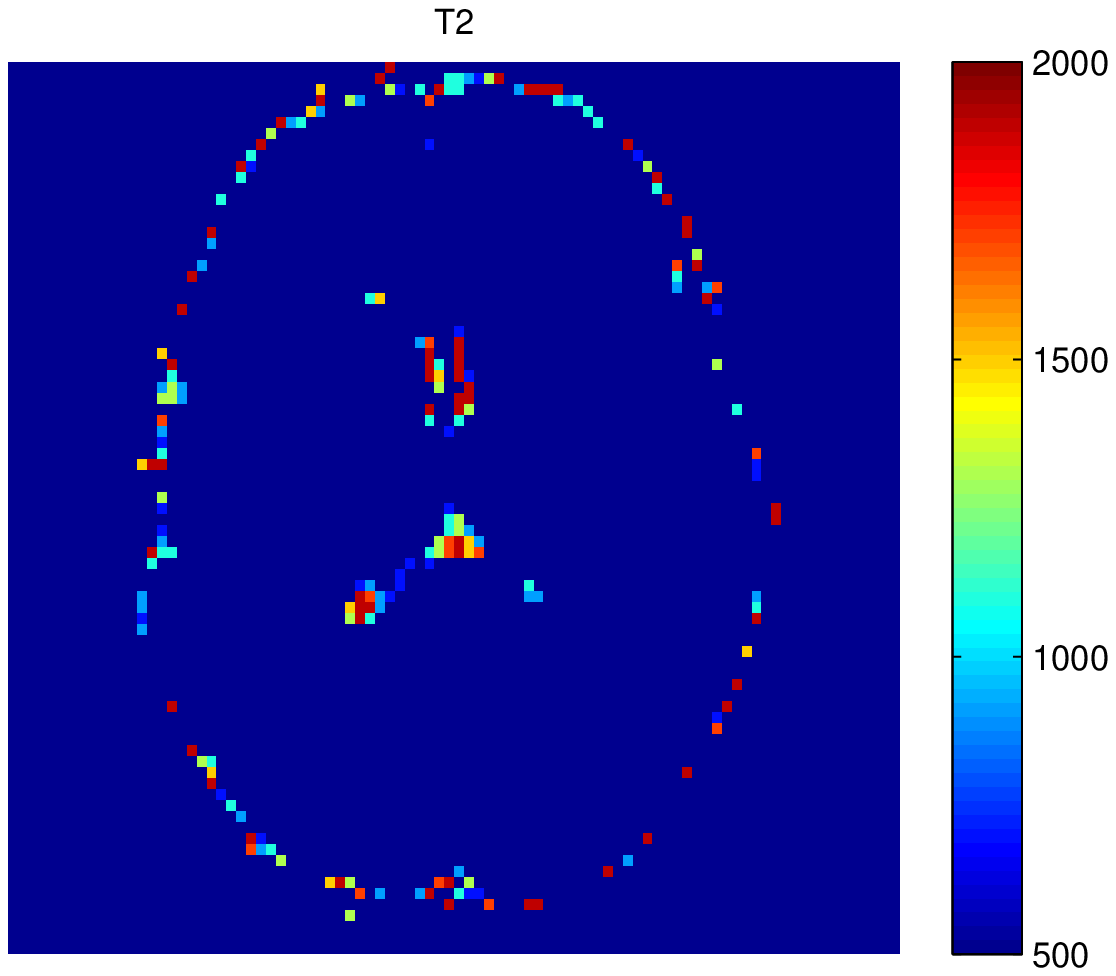}\vspace{-4mm}\\

  \caption{Comparison between T2 map reconstructions of BLIP, MBIR-MRF and FLOR with sequence length of 400 TRs. T2 color scales are above 500 milliseconds. It can be seen that FLOR provides T2 values for CSF that better match the literature values in this case.}
        \label{real_data_diff}
\end{figure*}

The results of T1, T2 and PD maps for BLIP, MBIR-MRF and FLOR appear in Fig.~\ref{real_data}. Since IR-bSSFP sequence has been used in this experiment, off-resonance frequency has also been computed and shown.  We used 109 different values in the range between -250 and 240 Hz. 
It can be seen that for T1, all iterative algorithms provide similar results, and T1 values of grey matter (GM), white matter (WM) and cerebrospinal fluid (CSF) regions correspond to similar values that appear in the literature (see Table 1 in Ma et al.\cite{ma2013magnetic}) and in Fig.~\ref{fig2_reference_MRF}. While T2 results exhibit visible differences between the various methods, WM and GM values for all methods correspond to values that appear in the literature. However, both BLIP and MBIR MRF exhibit T2 values for CSF that are lower than those reported in the literature. This can be seen in Fig.~\ref{real_data_diff}, where the color scale for T2 is adjusted to 500-2000ms (T2 values for CSF are around 2000ms). Shortened T2 values in CSF were also reported in the original MRF experiment with 1000 TRs (and were justified as out-of-plane flow in this 2D experiment). In our case, using the same acquired data, it can be seen in Fig.~\ref{real_data_diff} that FLOR provides CSF values that better correspond to literature values, when compared to the other methods.

\section{Discussion}
\label{sec:discussion}

\subsection{Relation to previous works}
Although works that exploit the low rank structure of MRF sequences have been published in the past by others \mbox{\cite{mcgivney2014svd,cline2016model,zhao2015model,zhao2016model,liao2016acceleration,doneva2017matrix,zhao2017improved,asslander2017low}} and also by us \cite{mazor2016low}, our solution is unique mainly in the combination of convex modelling and the ability to enable a solution with quantitative values that do not exist in the dictionary. Our solution is based on soft-thresholding the singular values \cite{cai2010singular}, which is mathematically justified in \mbox{Appendix A}. 


Moreover, we compare our algorithm to both CS-based and low-rank based methods for MRF and demonstrate superior results. While BLIP treats the original MRF problem as an $\ell_0$ optimization problem, FLOR first solves the relaxed problem of (\ref{eq6}) and only then uses MF to extract the magnetic parameters. It leads to some beneficial properties such as convergence guarantees, and the ability to use the acceleration step as described in \mbox{Algorithm 4}, which is also novel in MRF reconstruction methods.  


\subsection{Computational complexity} FLOR is divided into two main components: The first recovers the imaging contrasts, and the second extracts the parameter maps from the recovered contrasts. 
The computational burden of FLOR lies in the low-rank projection step, or specifically, in the SVD calculation. This step does not exist in BLIP nor the original MRF reconstruction. However, there are efficient fast techniques to calculate the SVD \cite{drmac2005new}, required by FLOR. Moreover, unlike BLIP, and other low rank based algorithms such as MBIR-MRF, FLOR does not require the pattern recognition calculation at every iteration. 
Another time consuming step that exists in all algorithms is the non uniform Fourier transform. By using the acceleration step, FLOR reduces significantly the number of iterations required for convergence and therefore saves computational cost.

In addition, while previous implementations of CS-based reconstruction algorithms mainly use the inverse NUFFT (iNUFFT) algorithm, in our retrospective experiemtns we use SPURS \cite{kiperwas2016spurs}. Based on our observations, SPURS provides improved image reconstruction with the same computational complexity compared to iNUFFT.

\section{Conclusions}
\label{sec:conclusions}

We presented FLOR, a method for high quality reconstruction of quantitative MRI data using MRF, by utilizing the low-rank property of MRF data. Due to the fact that we exploit low-rank on top of the well known sparsity of MRF in the dictionary matching domain, we are able to obtain high quality reconstruction from highly under-sampled data. Our method is based on a convex minimization problem, leading to a solution in the dictionary subspace that overcomes its quantization error.

We provide results that are comparable to fully sampled MRF, using only 5\% of the data in a simulation environment. In addition, comparison against CS-based and low-rank based methods for MRF shows the added value of our approach in generating quantitative maps with less artifacts. Our results also consist of real-data, in-vivo experiments that exhibit FLOR superiority also for realistic multi-coil data acquisition. Future work will examine more sophisticated patch wise recoveries.

\section*{Acknowledgements} 

\noindent The authors wish to thank the Tel Aviv center for brain functions at Tel Aviv Sourasky Medical Center for providing the data required for experiment 1. We also wish to thank Dan Ma and Prof. Mark Griswold for providing the real data used in their experiments. This work was supported by the
Ministry of Science, by the ISF I-CORE joint research center of the
Technion and the Weizmann Institute, and by the European Union\textquoteright s
Horizon 2020 research and innovation programme under grant agreement
No. 646804-ERC-COG-BNYQ, Assaf Tal acknowledges the support of the Monroy-Marks Career Development Fund, the Carolito Stiftung Fund, the Leona M. and Harry B. Helmsley Charitable Trust and the historic generosity of the Harold Perlman Family. The authors have no relevant conflicts of interest to disclose.

\appendix
\section{}
\label{sec:appendix}

The basic implementation of FLOR, as described in \mbox{Algorithm 4} in the paper, aims to solve the following optimization problem:


\begin{equation}
\underset{\mat{X}\in \mathbb{D}}{\text{argmin}}  \frac{1}{2}\underset{i}{\Sigma}\left\Vert \mathbf{Y}_{:,i}-F_u\{\mathbf{X}_{:,i}\}\right\Vert _{2}^{2}+\lambda||\mat{X}||_{*}
\label{eqa1}
\end{equation}

\noindent where $F_u$ is the partial Fourier transform operator, $\mat{X}$ has dimensions $N^2 \times L$ and $\mathbb{D}=\{\mat{X}:\mathcal{N}(\mat{X})\supseteq \mathcal{N}(\mat{D})\}$. 

FLOR solves (\ref{eqa1}) using the incremental proximal method \cite{sra2012optimization}, which treats problems of the form:
\begin{equation}
\underset{\mat{X}\in \mathbb{D}}{\text{argmin}} \ \{\Sigma_{i}^{m} F_{i}(\mat{X})\}
\label{eqa2}
\end{equation}
\noindent where $F_{i}(\mat{X})=f_{i}(\mat{X})+h_{i}(\mat{X})$. The function $f_{i}(\mat{X})$ is convex and non-differentiable, $h_{i}(\mat{X})$ is a convex function and $\mathbb{D}$ is a non-empty, closed, and convex subspace. The general step in solving (\ref{eqa2}) is given by [27, (4.12)-(4.13)]:

\begin{subequations}
\begin{equation}
\begin{aligned}
\mat{Z}^{k}=\text{P}_{\mathbb{D}}(\mat{X}^{k}-\mu_k g_{i_k})\\
\end{aligned}
\label{eqa31}
\end{equation}
\begin{equation}
\begin{aligned}
\mat{X}^{k+1}=\underset{\mat{X}\in \mathbb{D}}{\text{argmin}} \ f_{i_k}(\mat{X})+\frac{1}{2\mu_k}\|\mat{X}-\mat{Z}^{k}\|_F^{2}
\end{aligned}
\label{eqa32}
\end{equation}
\label{eqa33}
\end{subequations}

\noindent where $g_{i_k}\in \partial h_{i_k}(\mat{X}^{k}) $, $\mu_k$ is a positive step size, and $\text{P}_{\mathbb{D}}$ is the projection operator onto ${\mathbb{D}}$ defined as

\begin{equation}
\begin{aligned}
\text{P}_{\mathbb{D}}(\mat{X})&= \underset{\mat{Z}\in \mathbb{D}}{\text{argmin}}\|\mat{Z}-\mat{X}\|_{F}^2.
\end{aligned}
\label{eqa6}
\end{equation}

\noindent The optimization problem, defined in the update step of $\mat{X}^{k+1}$, is referred to as the proximal gradient calculation of the non-differentiable $f_{i_k}$, under the constraint $\mat{X}\in \mathbb{D}$. 

Our problem in (\ref{eqa1}) corresponds to $m=1$ in (\ref{eqa2}) and
\begin{equation}
\begin{aligned}
h(\mat{X})&=\frac{1}{2}\underset{i}{\Sigma}\left\Vert \mathbf{Y}_{:,i}-F_u\{\mathbf{X}_{:,i}\}\right\Vert _{2}^{2}=\frac{1}{2}\|\mat{Y}-F_{u}\{\mat{X}\}\|_{F}^{2}\\ 
f(\mat{X})&=\lambda\|\mat{X}\|_{*}.
\end{aligned}
\label{eqa4}
\end{equation}
\noindent Therefore,
\begin{equation}
\begin{aligned}
\partial h(\mat{X})&=F_{u}^{H}\{\mat{Y}-F_{u}\{\mat{X}\}\}, 
\end{aligned}
\label{eqa5}
\end{equation} 

\noindent and,
\begin{equation}
\begin{aligned}
\text{P}_{\mathbb{D}}(\mat{X})&=\mat{X}\mat{D}^{\dagger}\mat{D}=\mat{X}\mat{P}.
\end{aligned}
\label{eqa6}
\end{equation}

\noindent The solution of (\ref{eqa32}) for $f(\mat{X})=\lambda \|\mat{X}\|_*$ without the constraint $\mat{X}\in \mathbb{D}$, is the singular value soft-thresholding operator (SVT) \cite{cai2010singular} defined as:
\begin{equation}
\text{SVT}_{\mu_k\lambda}(\mat{Z}^{k})=\mat{U}_r[\mat{\Sigma}_r-\mu_k\lambda\mat{I}]_+\mat{V}_r^H.
\label{eqa11}
\end{equation}
\noindent Here $\mat{\Sigma}_r$ is a diagonal matrix with the non-zero singular values of $\mat{Z}^{k}$ on its diagonal, $\mat{U}_r$ and $\mat{V}_r$ are the $r$ left and right singular vectors of the SVD of $\mat{Z}^k$, associated with the $r$ non-zero singular values, and $[x]_+=\text{max}(0,x)$.
In our case, since $\mat{Z}^k\in \mathbb{D}$ (as follows from (\ref{eqa31})) and the SVT calculation keeps the operand in the same subspace, the constraint $\mat{X}\in \mathbb{D}$ can be omitted. Therefore, (\ref{eqa32}) reduces to 
\begin{equation}
\mat{X}^{k+1}=\mat{U}_r[\mat{\Sigma}_r-\mu_k\lambda\mat{I}]_+\mat{V}_r^H.
\label{eqa11}
\end{equation}


Combining (\ref{eqa5}), (\ref{eqa11}) and (\ref{eqa6}), the incremental subgradient-proximal method for solving (\ref{eqa1}) consists of two updates in each iteration:
\begin{subequations}
\begin{equation}
\mat{Z}^{k}=(\mat{X}^{k}-\mu_k{F_{u}^{H}\{\mat{Y}-F_{u}\{\mat{X}^k\}\}})\mat{P}
\label{eqa7}
\end{equation}
\begin{equation}
\mat{X}^{k+1}=\mat{U}_r[\mat{\Sigma}_r-\mu_k\lambda\mat{I}]_+\mat{V}_r^H.
\label{eqa8}
\end{equation}
\end{subequations}

\noindent This constitutes the core of Algorithms 4. In our framework, the step sizes are set to constant, $\mu_k=\mu$, and $\lambda$ is chosen experimentally.

\bibliography{low_rank_mrf_journal}

\end{document}